\documentclass{article}

\usepackage{arxiv}

\usepackage[utf8]{inputenc} 
\usepackage[T1]{fontenc}    
\usepackage{hyperref}       
\usepackage{url}            
\usepackage{booktabs}       
\usepackage{amsfonts}       
\usepackage{nicefrac}       
\usepackage{microtype}      
\usepackage{lipsum}
\usepackage{graphicx}
\usepackage{threeparttable}
\usepackage{amsmath}
\usepackage{natbib}
\usepackage[dvipsnames]{xcolor}
\usepackage{multicol}
\usepackage{multirow}
\usepackage{glossaries}
\usepackage{diffcoeff}
\usepackage{ulem}
\usepackage{lineno}
\usepackage{floatrow}
\usepackage[labelfont=bf]{caption}
\usepackage[all]{hypcap}    

\newfloatcommand{capbtabbox}{table}[][\FBwidth] 

\hypersetup{colorlinks=true, linkcolor={TealBlue}, filecolor=magenta, urlcolor={ForestGreen}, citecolor={magenta}}

\newcommand{\mnras}{\textit{Monthly Notices of the Royal Astronomical Society}}
\newcommand{\aap}{\textit{Astronomy and Astrophysics}}
\newcommand{\apjs}{\textit{Astrophysical Journal, Supplement}}
\newcommand{\apj}{\textit{Astrophysical Journal}}
\newcommand{\apjl}{\textit{Astrophysical Journal, Letters}}
\newcommand{\ssr}{\textit{Space Science Reviews}}

\newacronym{sed}{SED}{Spectral Energy Distribution}
\newacronym{ssc}{SSC}{Synchrotron Self Compton}
\newacronym{ec}{EC}{External Compton}
\newacronym{blr}{BLR}{Broad Line Region}
\newacronym{dt}{DT}{Dusty Torus}
\newacronym{ebl}{EBL}{Extra-galactic Background Light}
\newacronym{mjd}{MJD}{Modified Julian Date}
\newacronym{roi}{ROI}{Region of Interest}
\newacronym{nir}{NIR}{Near Infrared}
\newacronym{agn}{AGN}{Active Galactic Nuclei}
\newacronym{smbh}{SMBH}{Super Massive Black Hole}
\newacronym{ic}{IC}{inverse Compton}
\newacronym{fsrq}{FSRQ}{Flat Spectrum Radio Quasar}
\newacronym{pl}{PL}{Power Law}
\newacronym{lp}{LP}{Log-Parabola}
\newacronym{bpl}{BPL}{Broken Power Law}
\newacronym{ts}{TS}{Test Statistic}

\title{Study of Temporal and Spectral variability for Blazar PKS 1830-211 with Multi-Wavelength Data}

\author{
 Jayant Abhir \\
  Department of Physics\\
  IIT Kharagpur\\
  \And
  Raj Prince \\
  Center for Theoretical Physics\\
  Polish Academy of Sciences
  \And
  Jophin Joseph \\
  Department of Physics\\
  IIT Kharagpur\\
  \And
  Debanjan Bose \\
  Department of Astrophysics \& Cosmology\\
  S N Bose National Centre for Basic Sciences\\
  \texttt{debanjan.tifr@gmail.com}
  \And 
  Nayantara Gupta \\
  Department of Astronomy \& Astrophysics\\
  Raman Research Institute
}

\makeglossaries

\begin{document}
\maketitle

\begin{abstract}
A study of the gravitationally lensed blazar PKS 1830-211 was carried out using multi waveband data collected by \textit{Fermi}-LAT, \textit{Swift}-XRT and \textit{Swift}-UVOT telescopes between \acrshort{mjd} 58400 to \acrshort{mjd} 58800 (9 Oct 2018 to 13 Nov 2019). Flaring states were identified by analysing the $\gamma$-ray light curve. Simultaneous multi-waveband \gls{sed} were obtained for those flaring periods. A cross-correlation analysis of the multi-waveband data was carried out, which suggested a common origin of the $\gamma$-ray and X-ray emission. The broadband emission mechanism was studied by modelling the \gls{sed} using a leptonic model. Physical parameters of the blazar were estimated from the broadband \gls{sed} modelling. The blazar PKS 1830-211 is gravitationally lensed by at least two galaxies and has been extensively studied in the literature because of this property. The self-correlation of the $\gamma$-ray light curve was studied to identify the signature of lensing, but no conclusive evidence of correlation was found at the expected time delay of 26 days. 
\end{abstract}


\section{Introduction} \label{sec:intro}
Blazars belong to a class of \gls{agn} with one of the jets within a few degrees of the observer's line of sight \citep{Urry_1995}. They are known for their strong and stochastic variability in low energy radio to high energy $\gamma$-ray frequency. The observed emission is highly anisotropic and non-thermal. The physical processes involved in the broadband emissions are synchrotron and \gls{ic} scattering. Synchrotron process is responsible for the emission in radio to soft X-rays and \gls{ic} process is believed to up-scatter the low energy photons to high energy $\gamma$-rays in a leptonic scenario. Diverse time scales of variability ranging from minutes to several days have been observed in their emission and in some cases up to decades \citep{Goyal_2017, Goyal_2018, Goyal_2020}. The multi-wavelength studies of blazars suggest that the nature of the emission is very complex. The nature of variability in the various wavebands differs from source to source and sometimes from flare to flare for the same source. 

The observed broadband \gls{sed} of blazars is also very complex. As an example, the study by \citet{Prince_2020} found that the \gls{sed} modelling of 3C 279 using the one zone leptonic emission model during the flares in 2017-2018, favours the scenario where the $\gamma$-ray emission region is located at the outer boundary of the \gls{blr}. Another study of 3C 279 during the flare of March 2014 \citep{Paliya_2015} explains the $\gamma$-ray flaring by a one-zone leptonic scenario, where the seed photons from the \gls{blr} and the \gls{dt} are required. On the other hand, the $\gamma$-ray flare of December 2013 for the same source suggests that to explain the broadband emission, the \gls{sed} can be fitted by both lepto-hadronic and two-zone leptonic model \citep{Paliya_2016}. This clearly suggests that even for the same source, different physical processes are possible at different epochs, reflecting the complexities seen in the emission from blazar jets. 

PKS 1830-211 is a gravitationally lensed \gls{fsrq}, as identified in 3FGL catalog \citep{3fgl_catalog} (Catalog Identifier: 3FGL J1833.6-2103) at redshift $z$=2.507 \citep{1999ApJ...514L..57L}. It is also known as TXS 1830-210, RX J1833.6-210 or MRC 1830-211. It is one of the brightest high redshift \textit{Fermi} blazars and the brightest lensed object in X-ray and $\gamma$-ray energies \citep{2015ApJ...799..143A}.

A time delay of $26^{+4}_{-5}$ days was detected from the light curve of two lensed images taken with ATCA by \cite{1996ApJ...472L...5L}. The first evidence of gravitational lensing phenomena in the $\gamma$-ray light curve of PKS 1830-211 was detected between 2008 and 2010 \citep{2011A&A...528L...3B}. They reported a time delay of $27.5\pm 1.3$ days with 3$\sigma$ confidence level using double power spectrum analysis. The October 2010 $\gamma$-ray outburst and the second brightest flaring period from December 2010 to January 2011 have been studied by \citet{2015ApJ...799..143A}. They did not find a significant correlation between X-ray and $\gamma$-ray variability and no time delay in the $\gamma$-ray emission due to the lensing.

The complex nature of PKS 1830-211 is discussed in detail in \cite{2015ApJ...799..143A}. PKS 1830-211 is a compound lensing system with possible micro-lensing/milli-lensing substructures with two foreground lensing galaxies at redshift 0.886 and 0.19. There could be many possible reasons for non observation of time delay in the $\gamma$-ray emission in the study by \citet{2015ApJ...799..143A} including micro/milli-lensing effects, the need for a refined strong lensing model, emission being obscured by absorbing material in the jet or physical processes or variations from the geometric configuration of the jet.

The H.E.S.S. array observed this source \citep{2019MNRAS.486.3886H} in August 2014 following a flare alert by the \textit{Fermi}-LAT collaboration. More than 12 hours of good quality data with an analysis threshold of 67 GeV was analysed and a 99$\%$ confidence upper limit on the average high energy $\gamma$-ray flux was obtained in the energy range of 67 GeV and 1 TeV. The long term \textit{Fermi}-LAT light curve of the source has been studied in \citet{2020ApJS..250....1T}.

A strong flaring period between 2018 October and 2019 November has been studied in this work. A five-fold increase in the $\gamma$-ray flux has been observed. Three flares have been identified and modelled with time dependent leptonic model. The multi-waveband data analysis is discussed in section 2. The multi-waveband light curves are studied in section 3. The spectral energy distribution and its modelling are discussed in section 4. Our results are discussed in section 5.

\section{Multi-waveband observations and data analysis}
The following section describes the analysis of the \textit{Fermi}-LAT, \textit{Swift}-XRT and \textit{Swift}-UVOT data used to obtain the multi frequency light curves and identification of the flaring periods in Section \ref{sec:mw_lightcurve}. Simultaneous multi waveband \gls{sed} was also obtained using this data which has been modelled in Section \ref{sec:SED_modelling}.

\subsection{High energy \texorpdfstring{$\gamma$}{gamma}-ray observations of \textit{Fermi}-LAT}
The data collected by the \textit{Fermi} Large Area Telescope (LAT) instrument on-board the \textit{Fermi} $\gamma$-ray Space Telescope in the 100 MeV to 300 GeV energy range was used to obtain the $\gamma$-ray light curve and $\gamma$-ray \gls{sed} data points for PKS 1830-211 (4FGL J$1833.6-2103$, \citealt{Ajello_2020}). \textit{Fermi}-LAT is a pair conversion $\gamma$-ray detector in orbit since June 2008. It has a large FoV of about 2.4 sr and single photon resolution of $<3.5^{\circ}$ at 100 MeV energy and improves to $<0.6^\circ$ for energies greater than 1 GeV \citep{2009ApJ...697.1071A}. A likelihood analysis is required to identify potential high energy $\gamma$-ray sources in the sky within a \gls{roi} using an input model. The analysis was done using the \texttt{fermitools}\footnote{\href{https://github.com/fermi-lat/Fermitools-conda}{\texttt{fermitools}} GitHub repository} package {\texttt{v1.2.1}} (maintained by the \textit{Fermi}-LAT collaboration) and \texttt{python}. Limited use of the \texttt{fermipy}\footnote{\href{https://fermipy.readthedocs.io/en/latest/index.html}{\texttt{fermipy}} homepage} \citep{2017ICRC...35..824W} package was also made. 

The \textit{Fermi}-LAT data in the energy range of 0.1–300 GeV was collected from \acrshort{mjd} 58400 (9 Oct 2018) to \acrshort{mjd} 58800 (13 Nov 2019). Events were extracted from a circular \gls{roi} of $15^\circ$ centered around the source position (Right Ascension: 278.416, declination: -21.0611). Photon data files were filtered with ‘\texttt{evclass=128}’ and ‘\texttt{evtype=3}’ and the time intervals were restricted using ‘\texttt{(DATA\_QUAL>0)\&\&(LAT\_CONFIG==1)}’ as recommended by the \textit{Fermi}-LAT team in the \texttt{fermitools} documentation. A zenith angle cut of $90^\circ$ was applied to avoid the contamination of the data from Earth’s limb.

Light curves with multiple different time bin sizes were computed and analysed using a model containing 269 point sources from the 4FGL catalog \citep{Ajello_2020} which were within a $15^\circ$ \gls{roi} around PKS 1830-211. The modelling of the isotropic and diffuse background was done using ‘\texttt{iso\_P8R3\_SOURCE\_V2\_v1.txt}’ and ‘\texttt{gll\_iem\_v07.fits}’ respectively. The spectral shapes and initial parameters were set using the values published in the 4FGL catalog and a total of $\sim$11 to 13 parameters (depending on the spectral emission model chosen for PKS 1830-211) were kept free for the likelihood analysis including parameters for two sources which were within $3^\circ$ of PKS 1830-211 and the diffuse and isotropic background. The $\gamma$-ray spectral points were modeled with three different spectral models described below - \gls{pl}, \gls{bpl} and \gls{lp} : 
\begin{linenomath*}
\begin{align}
    \diff{N(E)}{E} &= N_0\times\left(\frac{E}{E_0}\right)^{-\alpha} &- \text{PL} \label{eq:power_law}\\
    \diff{N(E)}{E} &= \begin{cases} N_0\times\left(\frac{E}{E_0}\right)^{-\alpha_1} \,\, E < E_0\\
    N_0\times\left(\frac{E}{E_0}\right)^{-\alpha_2} \,\, E \geq E_0
    \end{cases} &- \text{BPL} \label{eq:broken_power_law}\\
    \diff{N(E)}{E} &= N_0\times\left(\frac{E}{E_0}\right)^{-(\alpha+\beta \, \text{log}(E/E_0))} &- \text{LP}
    \label{eq:log_parabola}
\end{align}
\end{linenomath*}
where $N_0$ is the pre-factor, $E_0$ is the energy scaling factor/break energy and $\alpha$, $\alpha_i$ are spectral indices. The parameter $\beta$ in \gls{lp} model is known as the curvature index. 

Unbinned likelihood analysis was performed for the source models with each of the 3 spectral emission models using \texttt{gtlike} (\citealt{1979ApJ...228..939C}, \citealt{1996ApJ...461..396M}). The light curve in Fig \ref{fig:combined_lightcurve} was produced using \gls{pl} model. It was found that the flux was sufficient that the relative error was not too high and \gls{ts} values were $> 20$ for as low as 6 hr binning of the light curve. 
Light curves with 12 hr, 1 day and 5 day bin sizes were also produced for a broader overview of the 400 day period and for the cross-correlation and self-correlation analysis described in Sections \ref{sec:cross_correlation} and \ref{sec:self_correlation}. 

\subsection{X-ray observations of \textit{Swift}-XRT}
The X-ray data was collected in the energy range 0.3-8.0 keV by the X-Ray Telescope (\textit{Swift}-XRT, \citealt{2004SPIE.5165..201B}) onboard the Neil Gehrels \textit{Swift} Observatory (launched in November 2004) and analyzed for five intervals including three flaring periods, one pre-flare interval and one post-flare interval identified using the $\gamma$-ray light curve (described in Section \ref{sec:mw_lightcurve}). In the time period from \acrshort{mjd} 58545 to 58665, \textit{Swift}-XRT recorded 38 observations for the source. Data was unavailable for the rest of the time period of \textit{Fermi}-LAT observations. The flares in X-ray were found to be almost simultaneous with \textit{Fermi}-LAT flares. 

The data was analyzed with \href{https://heasarc.gsfc.nasa.gov/docs/software/heasoft/}{HEAsoft} package (\texttt{v6.26.1}) and \href{https://heasarc.gsfc.nasa.gov/xanadu/xspec/}{\texttt{XSPEC}} (\texttt{v12.10.1f}). The \href{https://www.swift.ac.uk/analysis/xrt/xrtpipeline.php}{\texttt{xrtpipeline}} package was used to obtain clean event files from the raw data. The source and background events were selected from circular regions of radii 20 pixels (1 pixel $\sim2.36''$) and 40 pixels respectively. The tool \texttt{xselect} was used to extract the spectrum files from the cleaned event files. The spectrum file finally used in \texttt{XSPEC} for the modelling.

The redistribution matrix file (RMF) was obtained from the HEASARC calibration database\footnote{HEASARC \href{https://heasarc.gsfc.nasa.gov/docs/heasarc/caldb/caldb_supported_missions.html}{calibration database}} and the ancillary response files (ARF) were generated using the tool \href{https://heasarc.gsfc.nasa.gov/ftools/caldb/help/xrtmkarf.html}{\texttt{xrtmkarf}}. The source event file, background event file, response file and ancillary response files were tied together using the tool \href{https://heasarc.gsfc.nasa.gov/lheasoft/ftools/fhelp/grppha.txt}{\texttt{grppha}} and then binned together to obtain a minimum of 20 counts. Since the count rate is well below 0.5 counts $s^{-1}$ for all 38 observations, pileup corrections were deemed unnecessary. The individual observations were combined using \href{https://heasarc.gsfc.nasa.gov/ftools/caldb/help/addspec.txt}{\texttt{addspec}} to get the spectra for a particular period and used in generating the \glspl{sed} for the five intervals mentioned before. The data was fitted with a basic power-law, $F(E) = K E^{\Gamma_x}$ using \texttt{XSPEC}. For this fit, neutral hydrogen column density was fixed at $N_{H}=21.9 \times 10^{20} \text{cm}^{-2}$ \citep{2006A&A...453..829F}. The results of this power law fit can be seen in Table \ref{tab:AnalysisResult}. 

\subsection{UV-optical observation of \textit{Swift}-UVOT}
The \textit{Swift}-Ultra-Violet/Optical Telescope (\textit{Swift}-UVOT, \citealt{Roming2005}) made the UV-optical observations which were used in the multi waveband light curve and \gls{sed} modelling. The observations between \acrshort{mjd} 58545-58665 were used in the present work. \textit{Swift}-UVOT obtains the data in three optical filters (V, B and U bands) and three UV filters (W1, M2 and W2 bands). The \texttt{uvotsource} tool was used to extract the magnitude. The magnitude (Mag) obtained from \texttt{uvotsource} is corrected for the reddening and the galactic extinction (A$_{\lambda}$) by using E(B-V) = 0.397 (\citealt{Schlafly_2011}), and the A$_{\lambda}$ was estimated from \cite{Giommi2006} for all the UVOT filters. Further, the corrected magnitudes were converted to flux using Swift-UVOT zero points (\citealt{Breeveld_2011}) and the conversion factors (\citealt{Giommi2006}). For constructing the UVOT \gls{sed}, we have summed the images in the individual filters from different observations using the tool \texttt{uvotimsum}. 

Dust absorption was accounted for by using the formula $\text{Mag}_0 = \text{Mag} - A_\lambda$. 
\begin{linenomath*}
\begin{equation}
    \text{Mag}_0 = -2.5 \, \text{log}(F) - Z_{pt}
\end{equation}
\end{linenomath*}
where $Z_{pt}$ is Zero point magnitude and F is flux in the units of $erg\,cm^{-2}\,s^{-1}$/ \AA.
The fluxes during the three flares are mentioned in Table \ref{tab:AnalysisResult}. For each of the filters, images from different observations were added using \texttt{uvotimsum} tool. 

\begin{table*}
\begin{tabular}{ l l c c c l }
    \hline
     & Parameter & Flare A & Flare B & Flare C & units \\
    \hline
    \multicolumn{2}{l}{\textit{Fermi}-LAT} & & & &\\
     & Spectral Index ($\alpha$) & $-2.39 \pm 0.01$ & $-2.41 \pm 0.01$ & $-2.29 \pm 0.01$ & - \\
     & Flux (F$_{0.1-300 \ \text{GeV}}$) & $10.74 \pm 0.14$ & $12.94 \pm 0.16$ & $13.76 \pm 2.30$ & $10^{-6}\, \text{photon(s) cm}^{-2}\,s^{-1}$ \\
     & Prefactor ($N_0$) & $148.9 \pm 2.4$ & $182.4 \pm 3.1$ & $177.9 \pm 4.2$ & $10^{-9}\, \text{photon(s) cm}^{-2}\,s^{-1}\, \text{MeV}^{-1}$ \\
     & \gls{ts} & 12920 & 30319 & 19160 &  - \\
    \hline
    \multicolumn{2}{l}{\textit{Swift}-XRT} & & & &\\
     & $\Gamma_x$ & $4.03 \pm 0.75$ & $1.65 \pm 0.38$ & $3.18 \pm 0.70$ & $10^{-1}$ \\
     & $K$ & $4.98 \pm 0.41$ & $4.07 \pm 0.17$ & $5.64 \pm 0.43$ & $10^{-4}\, \text{photon(s) cm}^{-2} \, s^{-1}\,\text{keV}^{-1}$ \\
     & Flux (F$_{0.3-8\, \text{keV}}$) & $13.8$ & $16.1$ & $17.7$ & $10^{-12}\, \text{erg cm}^{-2} \, s^{-1}$ \\
    \hline
    \multicolumn{2}{l}{\textit{Swift}-UVOT} & & & &\\
     & v band Flux & $24.35 \pm 0.39$ & $20.25 \pm 0.32$ & $22.62 \pm 0.36$ & $10^{-12} \, \text{erg cm}^{-2} \, s^{-1}$ \\
     & b band Flux & $19.37 \pm 0.46$ & $17.35 \pm 0.35$ & $17.51 \pm 0.42$ & $10^{-12} \, \text{erg cm}^{-2} \, s^{-1}$ \\
     & u band Flux & $7.79 \pm 0.16$ & $7.31 \pm 0.15$ & $8.09 \pm 0.16$ & $10^{-12} \, \text{erg cm}^{-2} \, s^{-1}$ \\
     & w1 band Flux & $4.41 \pm 0.18$ & $3.22 \pm 0.09$ & $3.99 \pm 0.16$ & $10^{-12} \, \text{erg cm}^{-2} \, s^{-1}$ \\
     & m2 band Flux & $0.68 \pm 0.09$ & - & $0.61 \pm 0.09$ & $10^{-12} \, \text{erg cm}^{-2} \, s^{-1}$ \\
     & w2 band Flux & $2.01 \pm 0.10$ & $1.69 \pm 0.05$ & $1.26 \pm 0.09$ & $10^{-12} \, \text{erg cm}^{-2} \, s^{-1}$ \\
    \hline
    \end{tabular}
    \caption{Results from spectral analysis of $\textit{Fermi}$-LAT, $\textit{Swift}$-XRT and $\textit{Swift}$-UVOT data for flares A, B and C. The \textit{Fermi}-LAT data is for a power law model, described in Eq \ref{eq:power_law}.}
    \label{tab:AnalysisResult}
\end{table*}

\section{Multi-waveband light curve}\label{sec:mw_lightcurve}
\begin{figure*}
    \centering
    \includegraphics[width=\textwidth]{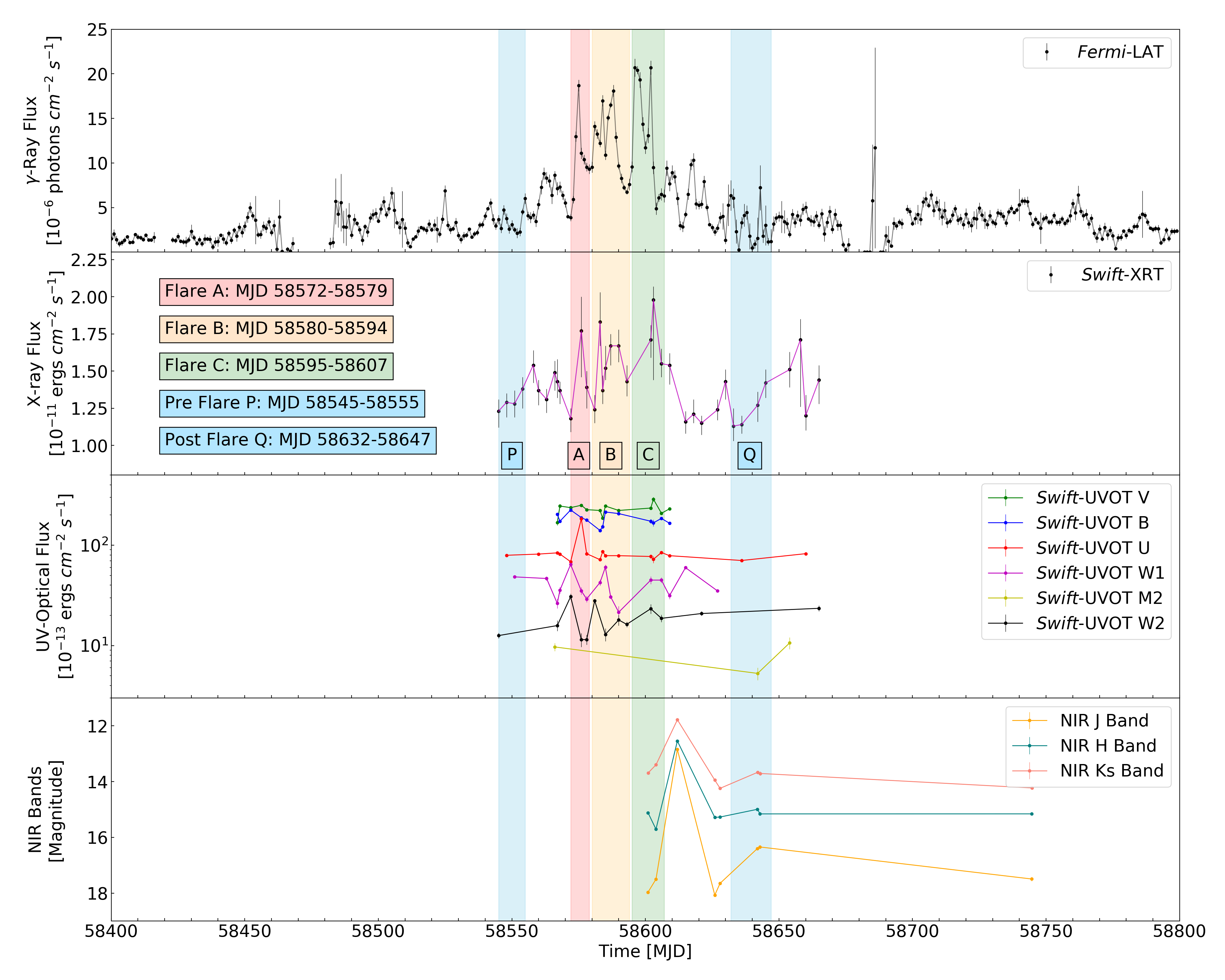}
    \caption{The multi waveband light curve for the full period from \acrshort{mjd} 58400-58800 ($9$ Oct $2018 - 13$ Nov $2019$)}. The $\gamma$-ray light curve is 1 day binned.
    \label{fig:combined_lightcurve}
\end{figure*}

The Multi-waveband light curve was produced by combining the data generated in the $\gamma$-ray, X-ray and UV-optical bands. The light curve can be seen in Fig \ref{fig:combined_lightcurve} with the various activity states marked in different color bands. Since the X-ray and UV-optical data is sparse, the choice of the flaring periods was aimed at maximising the number of data points within a single flaring period. The 400 day period was chosen based on the long term light curve of the source available on Fermi Space Science Center servers \footnote{PKS 1830-211 \href{https://fermi.gsfc.nasa.gov/ssc/data/access/lat/msl_lc/source/PKS_1830-211}{long term light curve}}. This period had significantly higher $\gamma$-ray flux, at least $10 \times$ compared to its flaring state studied in \cite{2015ApJ...799..143A}. The individual flares, Flare A, B and C and a pre-flare (P) and post-flare (Q) were then identified using the 6 hr and 1 day binned $\gamma$-ray data. \textit{Swift}-XRT and \textit{Swift}-UVOT observations were only available for a short period which limits the scope for the time period of the pre-flare and post-flare state since simultaneous data was required for meaningful constraints in the \gls{sed} modelling (Section \ref{sec:SED_modelling}). 

The X-ray light curve did not show as much variation as the $\gamma$-ray light curve and also contains a high state after the post-flare period (Q, marked in Fig \ref{fig:combined_lightcurve}) while no $\gamma$-ray or UV-optical counterpart was seen for this increase in X-ray flux. The \gls{nir} band data is just for reference and shows a high state (obtained via private communication with Luis Carrasco\footnote{ATeL $\#$\href{http://www.astronomerstelegram.org/?read=12784}{12784}}, ATeL $\#12784$). However, the \gls{nir} data has not been used for \gls{sed} modelling in Section \ref{sec:SED_modelling} since it is non-simultaneous for most of the relevant time periods. 

\subsection{Spectral model fits for the flares}
\begin{figure*}
    \centering
    \includegraphics[width=\textwidth]{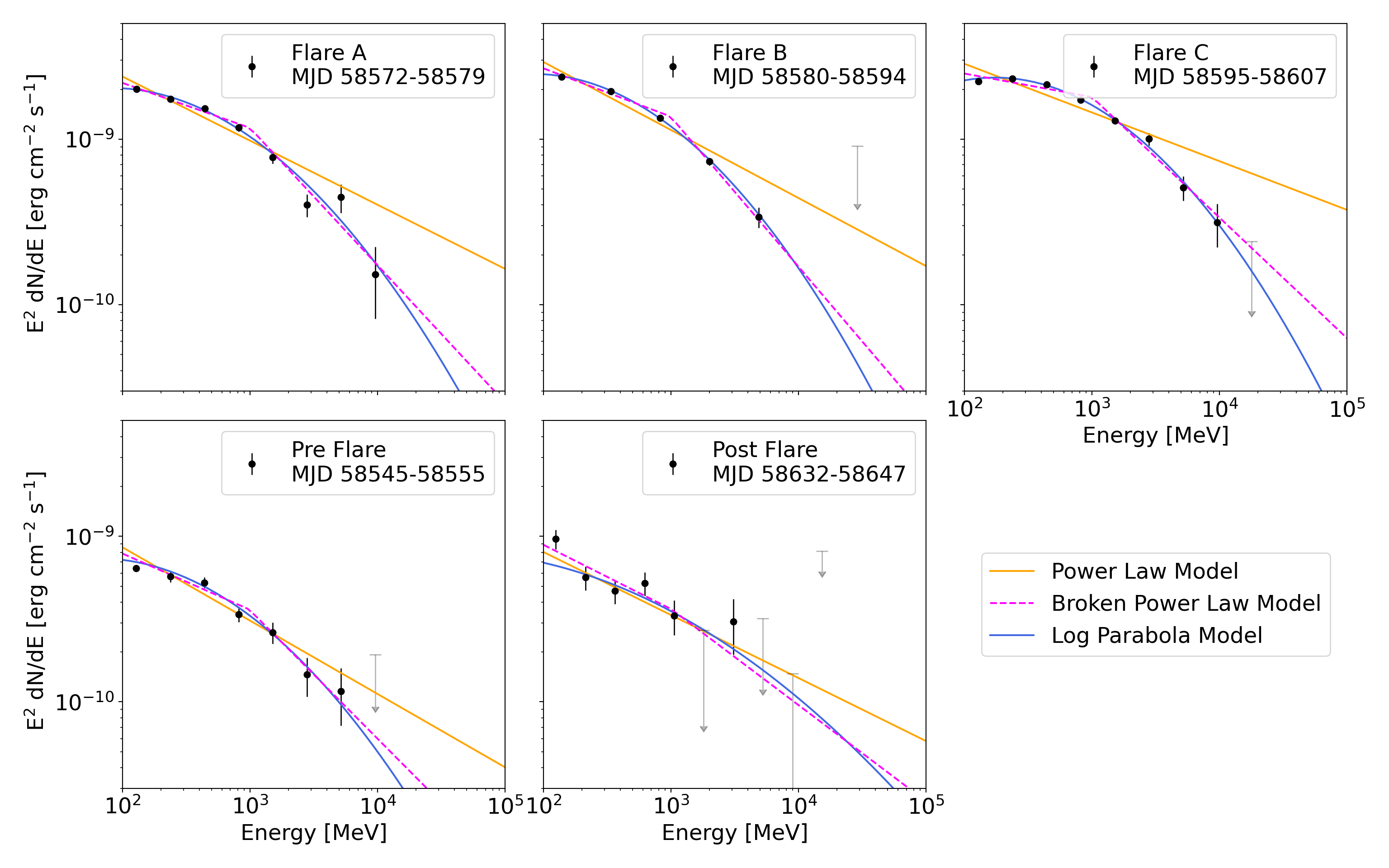}
    \caption{A comparison of the Log-parabola model with the power law model fitted to the $\gamma$-ray \gls{sed} . The Flaring states A, B and C are very similar in terms of the fit parameters and flux levels, while the pre-flare and post-flare states have much lower flux and are comparable to each other.}
    \label{fig:model_comparison}
\end{figure*}

The $\gamma$-ray spectral analysis was done following the standard procedure. The $\gamma$-ray \gls{sed} was fit with the 3 spectral models, \gls{pl}, \gls{bpl} and \gls{lp} described in Eq \ref{eq:power_law}, \ref{eq:broken_power_law} and \ref{eq:log_parabola}. It was found that \gls{bpl} and \gls{lp} models are statistically significant for the pre/post flare and flaring states of the source identified in Fig \ref{fig:combined_lightcurve} with about $4 \sigma$ significance compared to the power law model (estimated using the \gls{ts} values for the models). The comparison of the models can be seen in Fig \ref{fig:model_comparison}.

\subsection{Variability time scale computation}\label{sec:doubling_time}
\begin{figure}
    \centering
    \includegraphics[width=\columnwidth]{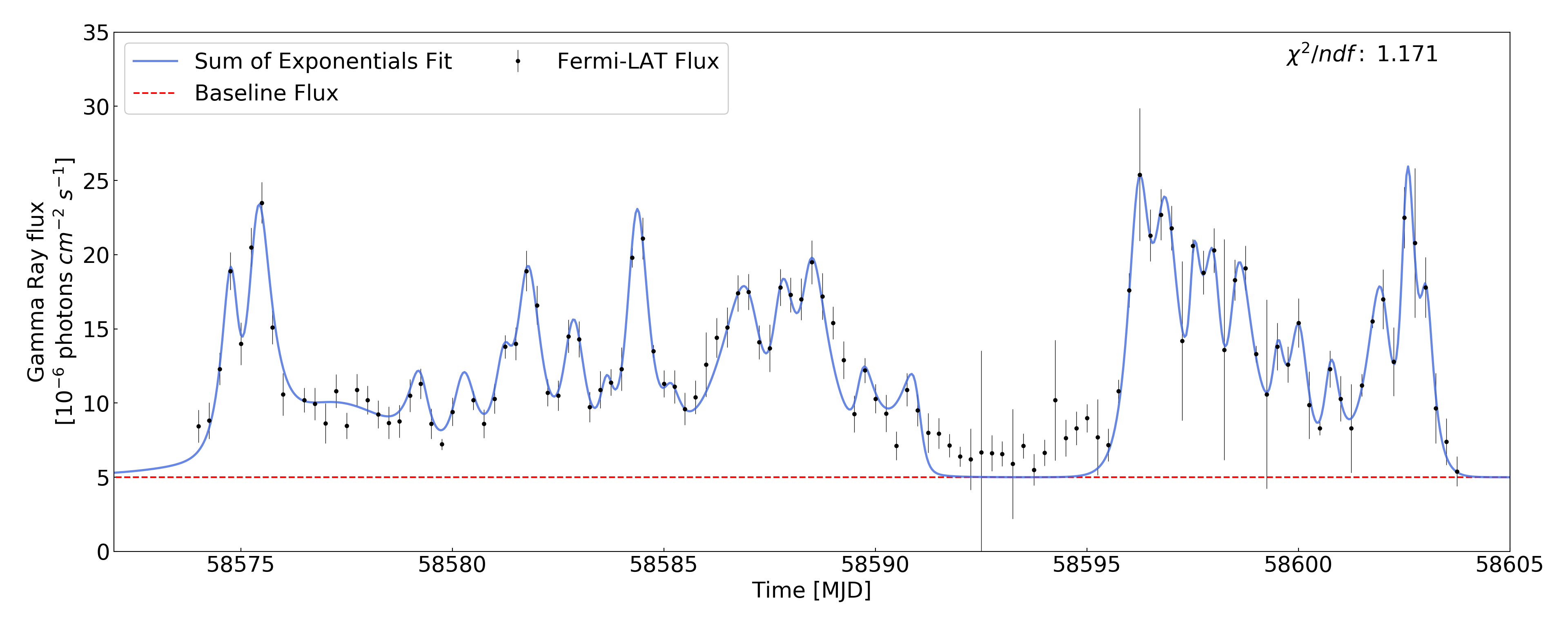}
    \caption{The 6 hr binned light curve during the flaring state, combined period for Flares A, B and C demarcated in Fig \ref{fig:combined_lightcurve} with a sum-of-exponentials fit for estimating rise and decay times $(T_R \text{ and } T_D)$.}
    \label{fig:6hr_fitted_lightcurve}
\end{figure}

The rise and decay time periods of peaks within the flaring period (Flare A, B, C in Fig \ref{fig:combined_lightcurve} combined, MJD 58570-58605) were calculated using a sum-of-exponentials fit. Fig \ref{fig:6hr_fitted_lightcurve} is the 6 hr binned $\gamma$-ray light curve, fitted with a sum-of-exponentials function of the form 
\begin{linenomath*}
\begin{equation}
    f(t) = a_0 + \sum_{i=1}^N 2 a_i \left[ \exp{\left(\frac{T_{i}-t}{T_{Ri}} \right)} + \exp{\left(\frac{(t-T_{i})}{T_{Di}}\right)} \right]^{-1}
    \label{eq:sum_of_exp}
\end{equation}
\end{linenomath*}
where $a_0$ is the baseline, $a_i$ are the scaling constants, $T_{Ri}$ and $T_{Di}$ are rise and decay time respectively of the peak ‘i’ and $T_i$ is a parameter approximately corresponding with the position of the peak maximum (exactly corresponds with the peak maximum for a symmetrical peak where $T_R = T_D$).

\begin{figure}
    \centering
    \begin{floatrow}
    \capbtabbox{
    \begin{tabular}{c c l}
    \hline
    T [hr] & $T_0$ [\acrshort{mjd}] & Rise/Decay\\
    \hline
    2.4 & 58589.65 & Decay \\
    3.0 & 58600.75 & Decay \\
    3.0 & 58574.80 & Rise \\
    3.6 & 58583.65 & Decay \\
    3.6 & 58598.00 & Rise \\
    \hline
    \vspace{7mm}
    \end{tabular}}
    {\captionsetup{width=.45\textwidth} 
    \captionof{table}{A few of the (fastest) rise and decay times for 6 hr fitted light curve during the flaring period from \acrshort{mjd} 58570-58610. Here, $T_0$ is the parameter $T_i$ in Eq \ref{eq:sum_of_exp} which only approximately corresponds to peak position when $T_R \neq T_D$ (asymmetric peaks).}
    \label{tab:rise_decay_times}}
    \hspace{2mm}
    \ffigbox{\includegraphics[width=1.1\columnwidth]{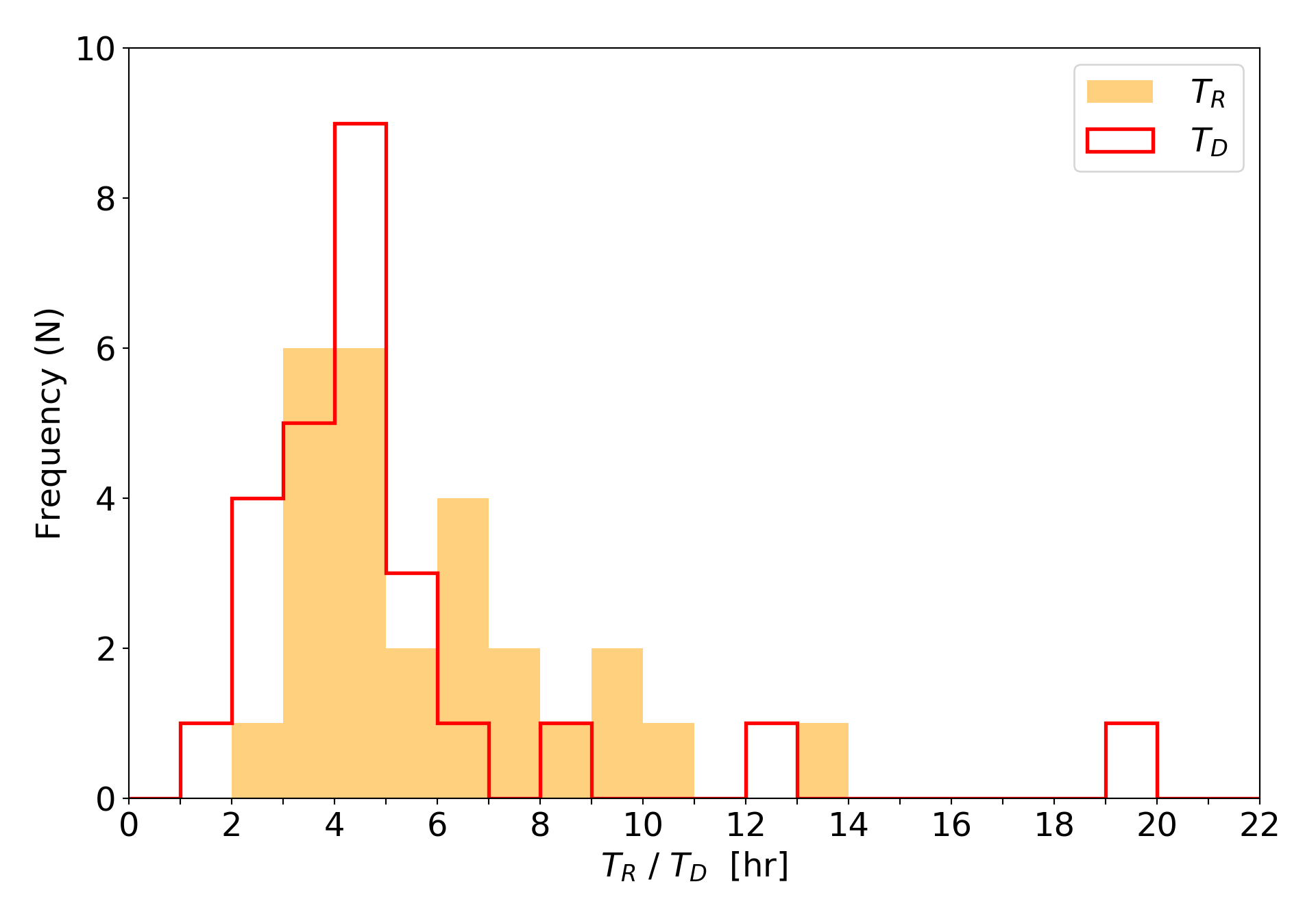}}
    {\captionof{figure}{The distribution of the rise and decay times during the flaring state based on the $T_R$ and $T_D$ parameters of all the peaks in the sum-of-exponentials fit.}
    \label{fig:rise_decay_histogram}}
    \end{floatrow}
\end{figure}

The fastest time period observed was 2.4 hr (see Table \ref{tab:rise_decay_times}) and distribution of $T_R$ and $T_D$ can be seen in Fig \ref{fig:rise_decay_histogram} with lesser spread in decay times. $T_D$ distribution is slightly faster, suggesting faster cooling time scale compared to energy injection time scale. The fastest rise and decay times are mentioned in Table \ref{tab:rise_decay_times} and their distribution is shown in Fig \ref{fig:rise_decay_histogram}.

The variability time scale, $t_{\text{var}}$ is a useful parameter that can set bounds to the size of the emission region. $t_{\text{var}} = ln(2) \times T_f$ where $T_f$ is the fastest decay/rise time \citep{2011ApJ...733L..26A}. The $T_R$ and $T_D$ values obtained using the sum-of-exponentials fit in Fig \ref{fig:6hr_fitted_lightcurve} suggests a $t_{\text{var}}$ of about 1.66 hr. The fitting was done using the 6 hr binned light curve, more detailed structure and shorter variability time scales were seen in 3 hr binned curve but the flux error increased significantly and \gls{ts} values reduced so we have only focused on the 6 hr (and higher) binned light curve data in this work.

\subsection{Flux-Index Correlation}
\begin{figure}
    \centering
    \begin{floatrow}
        \includegraphics[width=0.48\columnwidth]{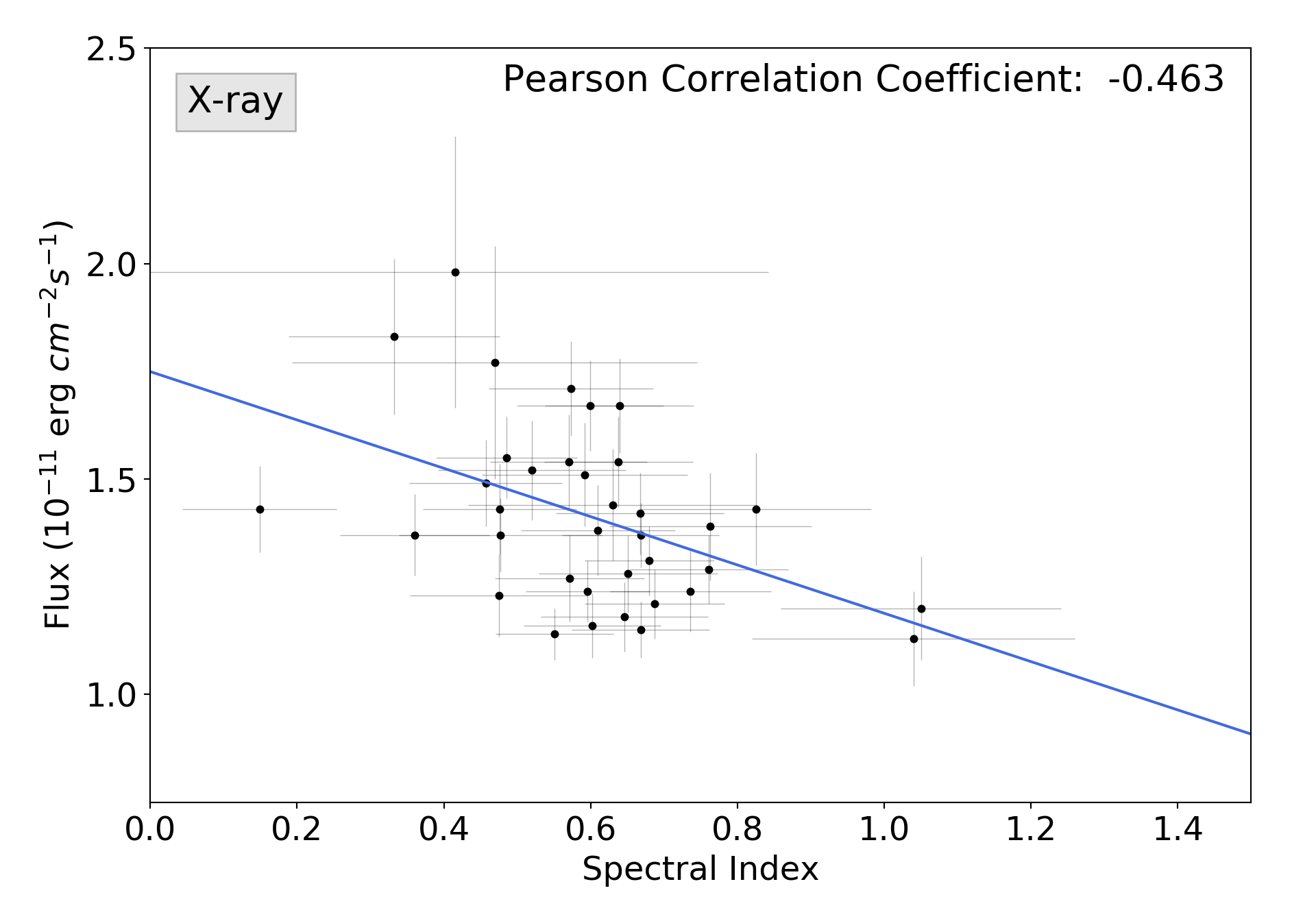}
        \hspace{2mm}
        \includegraphics[width=0.48\columnwidth]{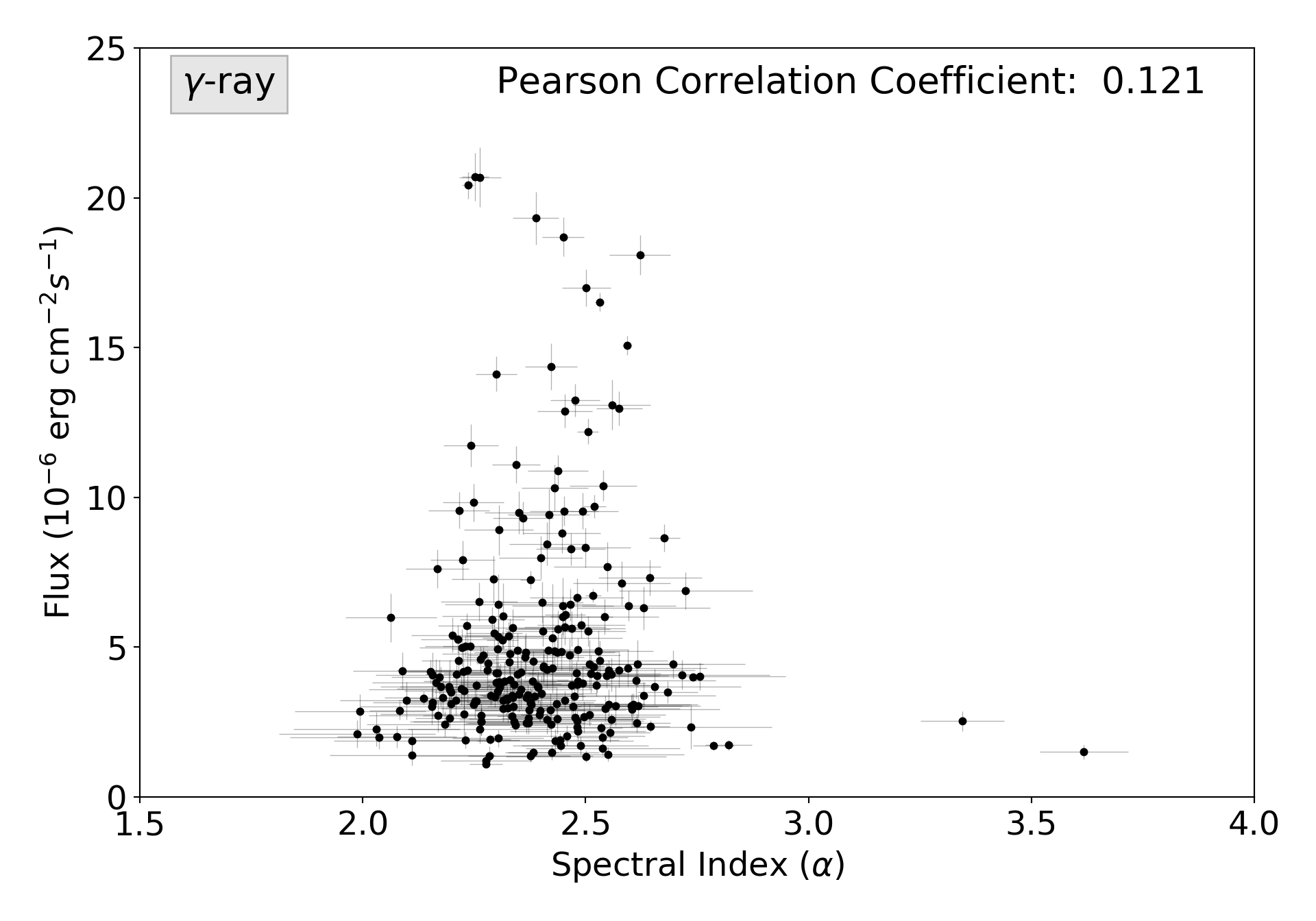}
    \end{floatrow}
    \caption{Flux vs Index plot for the X-ray data showing a decrease in the spectral index with increasing flux, the blue line is a linear fit for reference. The Flux vs Index plot for $\gamma$-rays did not show any meaningful trend. The $\gamma$-ray data is 1 day binned in this plot, with 6hr and 3hr data showing even greater scatter.}
    \label{fig:flux_index}
\end{figure}

Flux vs Index correlation was studied to detect index hardening/softening. No such correlation was found for the $\gamma$-ray flux in any of the 6 hr, 12 hr, 1 day and 5 day binned light curves, the absolute value of the Pearson Correlation Coefficient $(R)$ was $ < 0.12$ for all these cases, the scatter was very high in all cases and linear regression was not meaningful. On the other hand, the X-ray data displayed a reduction in the index with increasing flux. The Pearson correlation coefficient was found to be $R = -0.46$ with the p-value $<< \alpha$-level (significance threshold) ($p = 0.0045$, $\alpha = 0.05$). The trend follows a straight line with slope= $-0.56$. The flux-index correlation can be seen in Fig \ref{fig:flux_index}.

\subsection{Cross-correlation between different wavebands}\label{sec:cross_correlation}
Cross-correlation analysis between $\gamma$-ray, X-ray and UV-optical band light curves was carried out using the Discrete Correlation Function (DCF) as the data points are discrete and the sampling is uneven, especially in X-ray and UV-optical data. The DCF allows computation of a correlation coefficient without the use of interpolation for data sampled at different and/or variable rates. The DCF correlation coefficient is not bounded between $\pm 1$. The unbinned DCF function \citep{1988ApJ...333..646E} can be calculated for two data sets with data point ‘i’ in set 1 and data point ‘j’ in set 2 as:
\begin{linenomath*}
\begin{equation}
    \text{DCF}(\tau) = \sum_{i,j} \frac{\text{UDCF}_{ij}}{M}
\end{equation} 
\begin{equation}
    \text{UDCF}_{ij}  = \frac{(a_i - \overline{a})\,(b_j - \overline{b})}{\sqrt{(\sigma^2_a-e^2_a)\,(\sigma^2_b-e^2_b)}}
    \label{eq:dcf}
\end{equation}
\end{linenomath*}
where $\tau$ is the DCF bin size, $\Delta t_{ij}=(t_j-t_i)$ is the lag for the pair ($a_i,b_i$) and M is the total number of pairs for which \\$(\tau - \Delta\tau/2) \leq \Delta t_{ij} \leq (\tau + \Delta\tau/2)$. $\overline{a}$ and $\overline{b}$ are the averages of $a_i$ and $b_i$ respectively. $\sigma$ and $e$ are the standard deviation and measurement error associated with each set. The error in DCF can be calculated as:
\begin{equation}
    \sigma_{\text{DCF}(\tau)}=\frac{1}{M-1} \sqrt{\sum_{i,j}(\text{UDCF}_{ij} - \text{DCF}(\tau))^2} 
\end{equation}
A positive correlation coefficient implies that the first time series is leading with respect to the second time series and a negative coefficient implies the first series lagging behind the second. All subsequent discussion and plots have the $\gamma$-ray light curve as the first time series.

\begin{figure*}
    \centering
    \includegraphics[width=\textwidth]{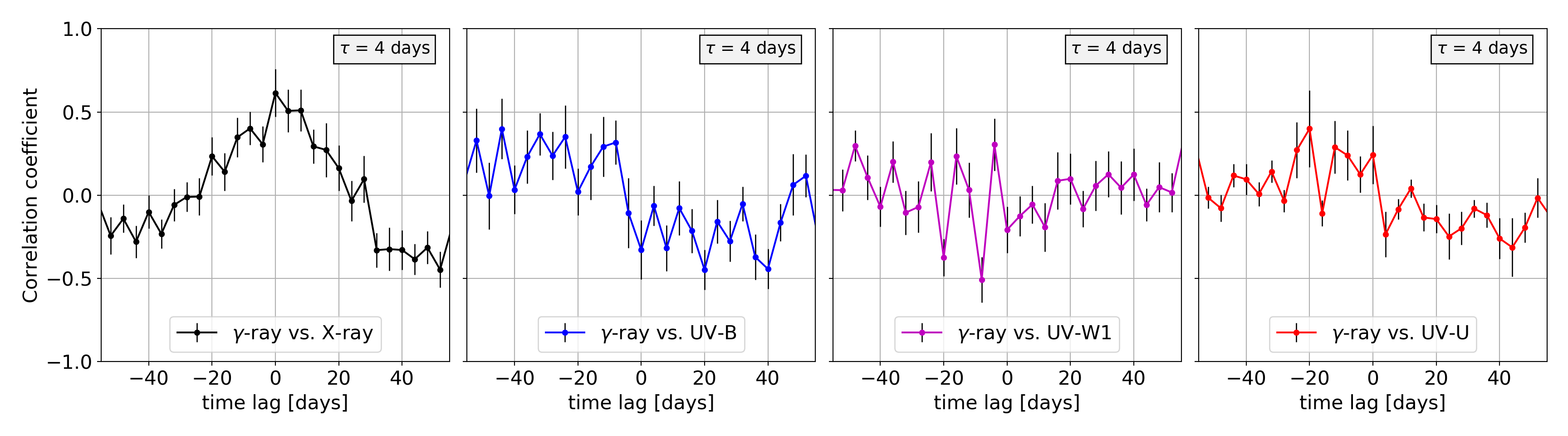}
    \caption{The DCF correlation coefficient between $\gamma$-ray and X-ray/UV-optical bands. The 1 day binned $\gamma$-ray light curve was used in the computation of cross correlation instead of the 6 hr binned light curve based on DCF bin size considerations. $\tau$ is the DCF bin size, kept at 4 days for all of the subplots (based on the average time resolution of X-ray and UV-optical light curves)}
    \label{fig:cross_correlation}
\end{figure*}

It was found that the $\gamma$-ray and X-ray flux is correlated with zero lag based on the maxima of the correlation coefficient plot (Fig \ref{fig:cross_correlation}) suggesting that the emission region for photons in these frequencies is co-spatial i.e. these photons originate in a single emission region. However, no such conclusions can be made for the $\gamma$-ray vs UV-optical band DCF, the position of the correlation coefficient peak varies for closely spaced UV-optical frequencies. This might be due to the fact that the UV-optical light curve has fewer peaks than the $\gamma$-ray light curve (Fig \ref{fig:combined_lightcurve}).

\subsection{Self-correlation and the effect of Gravitational Lensing in the \texorpdfstring{$\gamma$}{gamma}-ray light curve} \label{sec:self_correlation}
Since the blazar PKS 1830-211 is gravitationally lensed, it is expected to show some time delay in the self-correlation of the $\gamma$-ray light curve. Previous studies on PKS 1830-211 \citep{2011A&A...528L...3B} and other lensed blazars like B0218+357 \citep{2014ApJ...782L..14C} have detected self-correlation in the $\gamma$-ray light curve. The spatial resolution of \textit{Fermi}-LAT is not enough to resolve the lensed images which are within $1''$ of each other \citep{1988MNRAS.231..229P}, however, the time lag between different paths followed by photons that converge because of the lensing can be observed using self-correlation in the $\gamma$-ray light curve. The self-correlation was calculated using DCF in a similar procedure as described in Section \ref{sec:cross_correlation}. 

This effect of lensing for PKS 1830-211 has been studied extensively in the literature in multiple frequencies. There are 3 lensed images of the blazar \citep{2020A&A...641L...2M} but only 2 of them are bright, labelled as per their relative positions NE (North East) and SW (South West). A high resolution study of the blazar was done by \cite{1996ApJ...472L...5L} in radio frequencies where a delay of $26^{+4}_{-5}$ days was found between the 2 lensed images with NE being about $1.52 \times$ brighter than SW. The delay due to the lensing in $\gamma$-rays has been studied in detail in \cite{2011A&A...528L...3B} and \cite{2015ApJ...799..143A}. 

\begin{figure}
    \centering
    \includegraphics[width=0.6\columnwidth]{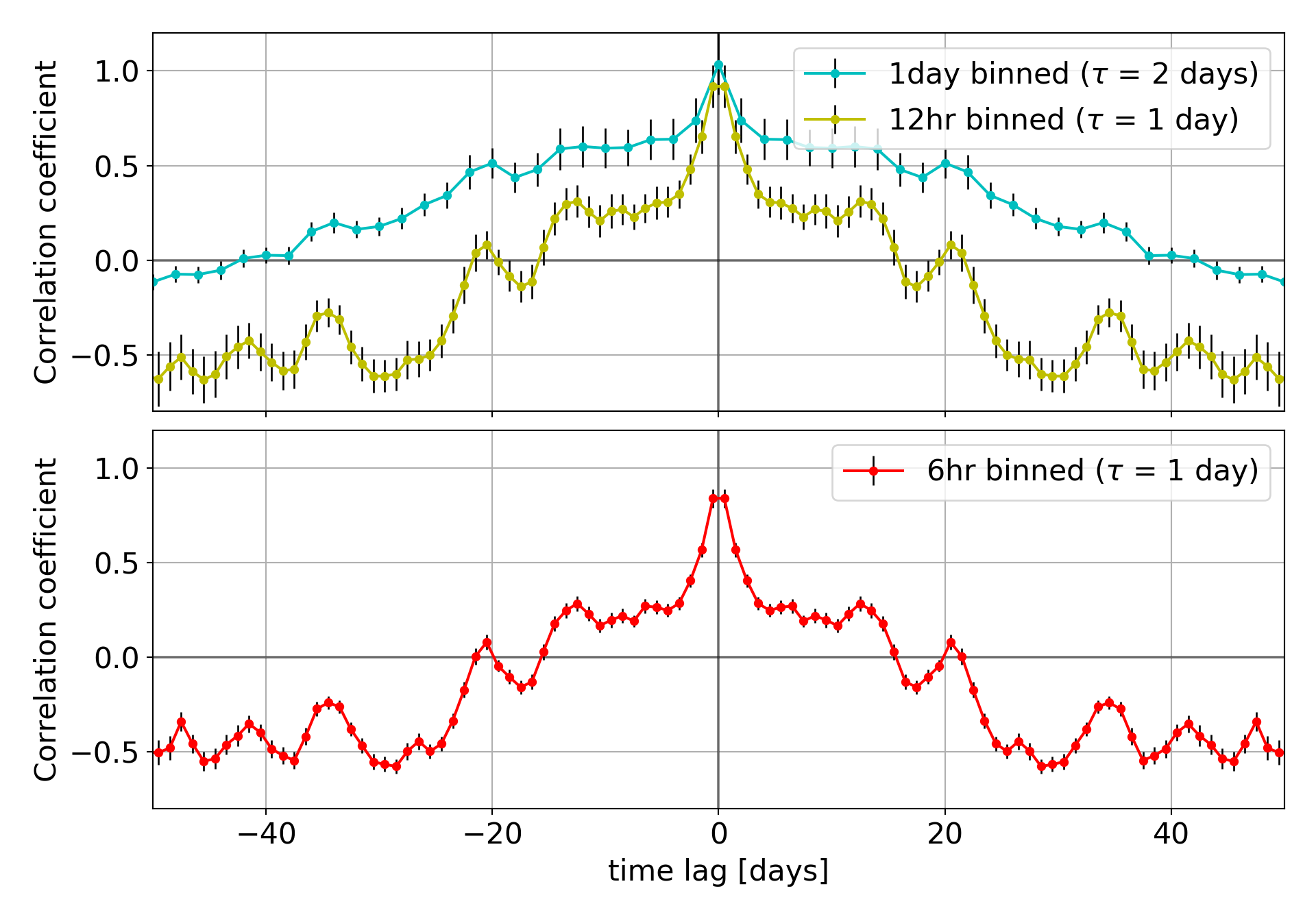}
    \caption{Self-correlation in the $\gamma$-ray light curve using DCF. $\tau$ is the DCF bin size, kept at $4 \times$ the sampling rate of the light curves. The x-axis and y-axis scales are the same for both the subplots.}
    \label{fig:self_correlation}
\end{figure}

We performed auto-correlation analysis on the 6 hr, 12 hr and 1 day binned $\gamma$-ray light curves using DCF but found no strong correlation at the expected delay introduced by lensing around 26 days as found in the radio bands by \cite{1996ApJ...472L...5L} or using the power spectrum in $\gamma$-rays by \cite{2011A&A...528L...3B}. We did see a similar structure of local maxima around 20 day lag/lead as observed in Figure 3 of \cite{2015ApJ...799..143A} for the 1 day and 12 hr binned data and to a lesser extent in the 6 hr binned data where a local maximum is visible but the correlation coefficient is almost zero. We also performed the Z-transformed DCF (zDCF) analysis \citep{2013arXiv1302.1508A} to make sure that the small value of correlation coefficient was not an issue of normalisation of the DCF coefficient but the zDCF results had similar correlation coefficient values as the DCF results and no new peaks could be identified. Our results are in line with the absence of a self-correlation peak at 26 days in the $\gamma$-ray light curve during the 2010 flaring studied by \cite{2015ApJ...799..143A}. The auto-correlation can be seen in Fig \ref{fig:self_correlation}.

\section{Spectral energy distribution and its modelling}\label{sec:SED_modelling}
Simultaneous multi-waveband \glspl{sed} were generated for the three flaring periods and the pre-flare and post-flare periods marked in Fig \ref{fig:combined_lightcurve} to study the emission mechanism. The model fitting was done using \texttt{GAMERA}\footnote{\href{http://libgamera.github.io/GAMERA/docs/main_page.html}{\texttt{GAMERA}} homepage}, a C++/\texttt{python} library for non-thermal emission modelling in $\gamma$-ray astronomy \citep{2015ICRC...34..917H}. 

A leptonic population with a log-parabola injection spectrum (Eq \ref{eq:log_parabola}) was used for the modelling. Leptonic models assume that relativistic leptons (mostly electrons and positrons) interact with the magnetic field in the emission region and produce synchrotron photons in the frequency region of the first hump of the \gls{sed}. The emission in the frequency region of the second hump of the \gls{sed} is reproduced by \gls{ic} scattering of a photon population further classified into \gls{ssc} or \gls{ec} categories based on the source of the seed photons. In the case of \gls{ssc} models \citep{1985A&A...146..204G, 1992ApJ...397L...5M} the seed photons for \gls{ic} process are the synchrotron photons produced by the same population of relativistic electrons. For the \gls{ec} models \citep{1993ApJ...416..458D, 1994ApJ...421..153S}, the seed photons can be sourced from one or more of the following external photon fields:
\begin{itemize}
    \item Direct emission in the optical region from the accretion disk
    \item Reprocessed emission in the UV-optical region from the \gls{blr}
    \item Reprocessed emission in the IR region from the \gls{dt}
\end{itemize}

For the \gls{sed} modelling the energy density of \gls{blr} in the comoving frame was estimated by \citep{2016A&A...594A.116H, Ghisellini2009}
\begin{linenomath*}
\begin{equation}
    U_\text{BLR}^{'} \sim \frac{\Gamma^2 \, \eta_\text{BLR} \, L_\text{disk}}{4 \pi \, c \, R^2_{\text{BLR}}}
    \label{eq:blr_density}
\end{equation}
\end{linenomath*}
where $\Gamma$ is the bulk Lorentz factor, $\eta_{\text{BLR}}$ is the fraction of disk emission processed in the \gls{blr}, typically around $10\%$, and $c$ is the speed of light in vacuum. R$_{\text{BLR}}$ is the radius of the \gls{blr}, and $L_\text{disk}$ is the accretion disk luminosity. Values for $R_{\text{BLR}} = 3.20 \times 10^{17} \, cm$ and $L_\text{disk} = 7 \times 10^{45} erg \, s^{-1}$ used in our calculations were taken from \cite{Celotti2008}.

\begin{figure*}
    \centering
    \includegraphics[width=\textwidth]{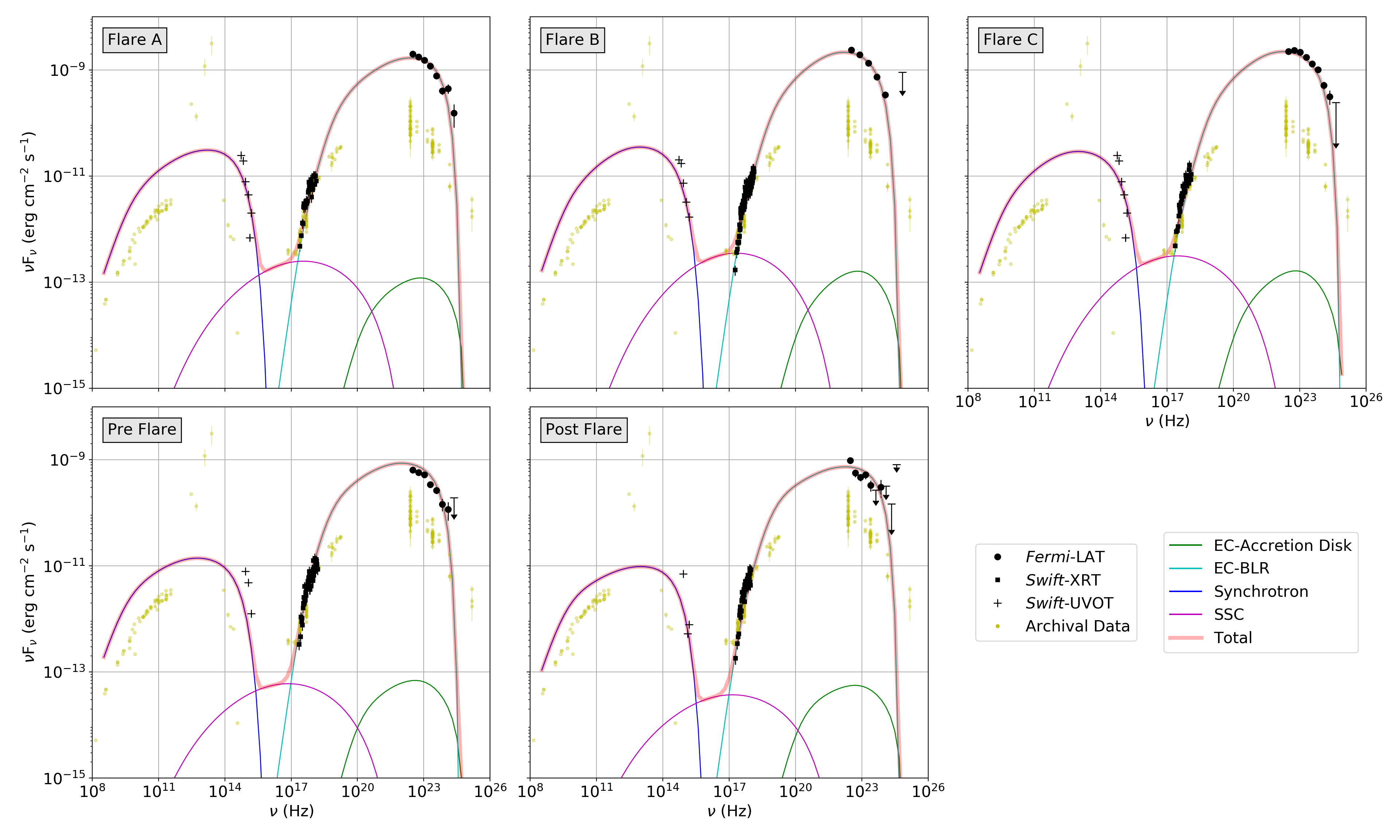}
    \caption{The multi frequency \gls{sed} data fit with a leptonic \gls{ec} model using \texttt{GAMERA}. \gls{ec}-\gls{blr} contribution is able to simultaneously explain the X-ray and $\gamma$-ray \gls{sed}. The scale and legend is the same for each subplot. Archival data in the background is for reference and comparison with the non-flaring state of the source. Our model included contribution from \gls{ic} scattering of CMB photons but the magnitude was too low compared to \gls{ec}-\gls{blr} to make any difference.}
    \label{fig:sed_model_fit}
\end{figure*}

If the emission region is within the \gls{blr} then the contribution of direct disk emission in IC scattering can not be ignored and the accretion disk energy density in the comoving frame can be derived using \citep{Dermer_2009}
\begin{linenomath*}
\begin{equation}
U'_{disk}=\frac{0.207 R_g \, l_\text{Edd} \, L_\text{Edd}}{\pi \, c \, d^3 \, \Gamma^2}
\end{equation}
\end{linenomath*}
where R$_g$, $l_\text{Edd} = L_\text{disk}/L_\text{Edd}$, and $d$ are the gravitational radius, the Eddington ratio and the location of the emission site from the \gls{smbh} respectively. The contribution from the accretion disk would be small as is seen in Figure \ref{fig:sed_model_fit}. Considering a \gls{smbh} mass of $\sim$10$^{9}$ M$_{\odot}$ \citep{2015ApJ...799..143A}, the gravitational radius is estimated as, R$_{g} \sim  1.5 \times 10^{14} \, cm$.

Our modelling of the blazar PKS 1830-211 during its flaring state is based on \gls{ec} from the \gls{blr} and the accretion disk. However, the contribution from the \gls{blr} dominates the \gls{ec} as also seen in Figure \ref{fig:sed_model_fit}. The UV-optical region data points constrain the synchrotron emission parameters in the model. As observed in \cite{2015ApJ...799..143A}, \gls{ssc} emission is too broad to adequately explain both the $\gamma$-ray and X-ray spectrum on its own. \cite{2015ApJ...799..143A} have modelled the X-ray and $\gamma$-ray frequencies with a combination of \gls{ssc} and \gls{ec}-\gls{dt}, while \cite{Celotti2008} have modelled PKS 1830-211 with \gls{ec}-\gls{blr}. Since a single \gls{ec} component from the \gls{blr} is able to fit the data in X-ray and $\gamma$-ray frequencies and the light curves for these frequencies have maximum correlation at zero lag, we come to the conclusion that the X-ray emission originates from the same electron population as the $\gamma$-ray emission, similar to \cite{2015ApJ...799..143A} who concluded that the X-ray emission originates from the low energy tail of the lepton population producing the $\gamma$-ray emission based on their \gls{sed} model but found no correlated variability.

The broadband \gls{sed} fit can be found in Fig \ref{fig:sed_model_fit} and the fitted parameters can be found in Table \ref{tab:sed_parameters}. We did not correct our \gls{sed} data points for lensing magnification following \cite{Celotti2008} and were able to satisfactorily model the \gls{sed} with good correspondence between the observed data and the model fits. The \gls{sed} data was below the threshold of 20 GeV for any significant \gls{ebl} correction (although we did observe photons up to 24.3 GeV, the flux at energies above 10 GeV was low and we only have upper limits beyond it, as can be seen in Fig \ref{fig:sed_model_fit}) using the \cite{2008A&A...487..837F} \gls{ebl} model, so no \gls{ebl} correction was applied. 

The \gls{sed} plots (Fig \ref{fig:sed_model_fit}) show archival data from various ground-based and space-based missions in a non-simultaneous period and does not affect the validity of our \gls{sed} modelling results. Our pre-flare and post-flare states have higher $\gamma$-ray flux compared to some other time periods (for example near MJD 58420) and higher flux in UV and X-ray bands compared to the archival data, the reason being the shorter period of \textit{Swift}-XRT and \textit{Swift}-UVOT observations compared to the $\gamma$-ray data. We prioritised the availability of simultaneous multi-waveband data while choosing flaring and non-flaring periods for the \gls{sed} modelling.

\subsection{Jet Power}
We have estimated the power carried by individual components (leptons, protons, and magnetic field) and the total jet power. The total power of the jet was estimated using 
\begin{linenomath*}
\begin{equation}
	 P_{\text{jet}}= \pi R_e^2 \, \Gamma^2 \, c (U^\prime_e+U^\prime_B+U^\prime_p)
	 \label{equ:Total_Power}
\end{equation}
\end{linenomath*}	
where $\Gamma$ is the bulk Lorentz factor; $U^\prime_B$, $U^\prime_\text{e}$ and $U^\prime_\text{p}$ are the energy density of the Magnetic field, electrons (and positrons) and cold protons respectively in the co-moving jet frame (primed quantities are in the co-moving jet frame while unprimed quantities are in the observer frame). The power carried by the leptons is given by
\begin{linenomath*}
\begin{equation}
	P_{\text{e}}= \frac{3\Gamma^2 \, c}{4 R_e} \int_{E_{\text{min}}}^{E_{\text{max}}} E \, Q(E) \, dE 
\end{equation}
\end{linenomath*}
where, $Q(E)$ is the injected particle spectrum. The integration limits, $E_{\text{min}}$ and $E_{\text{max}}$ are calculated by multiplying the minimum and maximum Lorentz factor ($\gamma_{min}$ and $\gamma_{max}$) of the electrons with the rest-mass energy of electron respectively.

The power due to the magnetic field was calculated using
\begin{equation}
    P_B= R_e^2 \, \Gamma^2 \, c \frac{B^2}{8}
\end{equation}
where $B$ is the magnetic field strength used to model the \gls{sed}. To calculate $U^\prime_p$ , we have assumed the ratio of the electron-positron pair to the proton is 10:1. We always maintain the charge neutrality condition. By calculating the total energy carried by protons and volume of the emission region, we computed total jet-power due to protons.

The Eddington luminosity for the source was estimated to be about $1.3 \times 10^{47} \text{erg} \, s^{-1}$ using a SMBH mass of $10^9 M_\odot$ \citep{2015ApJ...799..143A}. The calculated total power ($P_{\text{jet}}$) was below the Eddington limit for all cases and the power carried by individual components is mentioned in Table \ref{tab:sed_parameters}.

\begin{table*}
    \centering
    \begin{tabular}{l l l c c c c c}
        \hline
        Injected Lepton spectrum & & & \multicolumn{5}{c}{$\diff{N(E)}{E}=N_0\times\left(\frac{E}{E_0}\right)^{-(\alpha+\beta \, \text{log}(E/E_0))}$}\\
        \hline
        Parameter & Symbol & Unit & \multicolumn{5}{c}{Blazar State} \\
        \hline
        Fixed parameters & & & & & & & \\
        \hline
        Red shift & $z$ & - & \multicolumn{5}{c}{$2.507$} \\
        Distance & & parsec & \multicolumn{5}{c}{$10.58 \times 10^9$}\\
        \gls{smbh} Mass & & $M_\odot$ & \multicolumn{5}{c}{$10^9$} \\
        Doppler factor & $\delta$ & - & \multicolumn{5}{c}{$35$} \\
        Radius of Emission Region & $R_e$ & $cm$ & \multicolumn{5}{c}{$1.2 \times 10^{17} $} \\
        Radius of \gls{blr} & $R_{\gls{blr}}$ & $cm$ & \multicolumn{5}{c}{$3.2 \times 10^{17} $} \\
        Location of emission region & $d$ & $cm$ & \multicolumn{5}{c}{$1.5 \times 10^{17} $} \\
        (along Jet axis) & & & \multicolumn{5}{c}{ }\\
        Accretion Disk Temperature & $T_{\text{Disk}}$ & $K$ & \multicolumn{5}{c}{$2 \times 10^6$} \\
        Accretion Disk Energy Density & $U'_{\text{Disk}}$ & erg $cm^{-3}$ & \multicolumn{5}{c}{$1.02 \times 10^{-3}$} \\
        \gls{blr} Temperature & $T_{\text{\gls{blr}}}$ & $K$ & \multicolumn{5}{c}{$1 \times 10^5$} \\
        \gls{blr} Energy Density & $U'_{\text{\gls{blr}}}$ & erg $cm^{-3}$ & \multicolumn{5}{c}{$9.0$} \\
        \hline
        Variable Parameters & & & Flare A & Flare B & Flare C & Pre-Flare & Post-Flare \\
        \hline
        Energy Scaling Factor & $E_0$ & MeV & 120 & 120 & 120 & 120 & 120 \\
        Spectral Index & $\alpha$ & - & 1.7 & 1.8 & 1.8 & 1.9 & 1.8 \\
        Curvature Parameter & $\beta$ & - & 0.3 & 0.5 & 0.5 & 0.3 & 0.2 \\
        Magnetic Field & $B$ & Gauss &  2.0 & 1.9 & 1.7 & 1.9 & 1.7 \\
        $\gamma_{\text{min}}$ & - & - & 9 & 9 & 9 & 3 & 9 \\
        $\gamma_{\text{max}}$ & - & - & 1400 & 1400 & 1800 & 1200 & 1000 \\
        Luminosity scale factor & $L_0$ & $\times 10^{49}$ & 180 & 250 & 250 & 100 & 80 \\ 
        \hline
        Power of the jet & & & Flare A & Flare B & Flare C & Pre-Flare & Post-Flare \\
        \hline 
        Power in the Magnetic field & $P_B$ & $10^{45}$ erg $s^{-1}$ & 86.40 & 77.97 & 62.42 & 77.97 & 62.42\\
        Power in Leptons & $P_\text{e}$ & $10^{45}$ erg $s^{-1}$ & 9.09 & 10.41 & 10.67 & 4.91 & 4.53 \\
        Power in protons & $P_\text{p}$ & $10^{45}$ erg $s^{-1}$ & 5.38 & 5.59 & 5.60 & 4.08 & 3.91 \\
        Total Power & $P_{\text{jet}}$ & $10^{45}$ erg $s^{-1}$ & 100.87 & 93.98 & 78.70 & 86.96 & 70.87 \\
        \hline
    \end{tabular}
    \caption{The list of parameters and their values for the \texttt{GAMERA} model in Fig \ref{fig:sed_model_fit}. The values of the fixed parameters were the same for all the flaring and non-flaring states, the values mentioned in this Table were chosen after initial coarse fit for the \gls{sed}. After that, the model fit was progressively improved by varying the parameters associated with the injected lepton population. The variable parameters $E_0$, $\alpha$ and $\beta$ correspond to the log-parabola spectral shape of injected leptons that interact with the various photon fields in the blazar to produce the output \gls{sed}. The Flares and Non-flaring states have very similar values of these parameters suggesting a single emission mechanism is in play.}
    \label{tab:sed_parameters}
\end{table*}

\section{Results and discussion}

\subsection{The emission region} \label{sec:emission_region}
The comoving size of the emission region $R'$ can be estimated from the variability time scale, $R' = c \, \delta \, t_{\text{var}}/(1+z)$ \citep{2011ApJ...733L..26A}, where $t_{\text{var}} = 1.66$ hr, as computed in Section \ref{sec:doubling_time}. The value for $R'$ was estimated to be $1.78 \times 10^{15}$ cm, using a $\delta = 35$ found in our \gls{sed} modelling (Table \ref{tab:sed_parameters}). In the \gls{sed} modelling, we found the radius of the emission region, $R_e$ to be $1.2 \times 10^{17}$ cm however the values estimated using variability and the model are not necessarily in disagreement since the variability can be because of changes within a small volume of the emission region. 

The location of the emission region along the jet axis from the central \gls{smbh} can also be estimated from the variability time assuming a spherical emission region by using the expression $d \sim c \, \delta^2 \, t_{\text{var}} / (1+z)$ \citep{2011ApJ...733L..26A}. For $\delta = 35$, it was found to be $\sim 1.5 \times 10^{17} \, cm$ which puts the emission region within the \gls{blr} ($R_{\text{BLR}} = 3.2 \times 10^{17} \, cm$, \citealt{Celotti2008}), hence we have used a single zone \gls{ec}-\gls{blr} model while fitting the broadband \gls{sed}.

\subsection{High energy photons}
The total 400 day period was scanned to find the highest energy photons with greater than $90 \%$ probability of being associated with the source, with results mentioned in Table \ref{tab:highest_energy_photons}. Although the flux increased five-fold during the flaring period, the peak energy of photons did not go up significantly. Most of the highest energy photons were from the flaring period between \acrshort{mjd} 58570-58605, with some high energy $\gamma$-ray photons detected around \acrshort{mjd} 58760 although no flaring was observed at this time (as can be seen in Fig \ref{fig:combined_lightcurve}). The highest energy photon was found to be $24.3$ GeV, while many other blazars have been found to be the source of $30-50$ GeV photons (for example, PKS 1424-418 in \cite{2021MNRAS.501.2504A}, although at a lower redshift). This suggests the following two possibilities: 
\begin{itemize}
    \item The blazar is at $z=2.507$ which increases the probability of photon-photon interactions with the \gls{ebl} photons especially for the highest energy $\gamma$-ray photons. We found that no EBL correction was required for our data based on \cite{2008A&A...487..837F} \gls{ebl} model since our \gls{sed} data points were below 20 GeV but this does not exclude the possibility of a small number of very high energy photons being emitted by the source. 
    \item The emission region might be within the \gls{blr} or is being occluded by it. Our \gls{sed} model parameters and the variability time scale support this possibility and suggest that the emission region is within the \gls{blr}.
\end{itemize}

\begin{table}
    \centering
    \begin{tabular}{c c c}
        \hline
        Photon Energy & Arrival day [\acrshort{mjd}] & Probability \\
        \hline
        24.3 GeV & 58576.49 & 94.5$\%$ \\
        22.3 GeV & 58588.01 & 93.5$\%$ \\
        21.1 GeV & 58760.10 & 99.5$\%$ \\
        19.2 GeV & 58759.44 & 93.8$\%$ \\
        17.3 GeV & 58569.70 & 90.7$\%$ \\
        16.8 GeV & 58581.34 & 98.4$\%$ \\
        16.1 GeV & 58741.85 & 99.3$\%$ \\
        16.0 GeV & 58580.67 & 95.2$\%$ \\
        15.2 GeV & 58594.10 & 98.4$\%$ \\
        \hline
    \end{tabular}
    \caption{High energy $\gamma$-ray photons with Energy $> 15$ GeV detected from the source during the 400 day period from \acrshort{mjd} 58400 to 58800. The probability is for the photon being associated with the source which is slightly lower than those for an isolated source since PKS 1830-211 has diffuse sources within the $15^\circ$ \gls{roi} used in the analysis.}
    \label{tab:highest_energy_photons}
\end{table}

\subsection{Minimum Doppler factor} \label{sec:min_doppler}
The minimum value of the Doppler factor, $\delta_{\text{min}}$ can be estimated for the flaring states based on $\gamma \gamma$ opacity arguments and the highest energy photon observed \citep{2009ApJ...697.1071A}:
\begin{linenomath*}
\begin{equation}
    \delta_{\text{min}} \sim \left[ \frac{\sigma_\text{T} \, D_\text{L}^2 \, (1+z)^2 \, \varepsilon \, F_{\text{X-ray}} }{4 \, t_\text{var} \, m_e \, c^4}\right]^{1/6}
\end{equation}
\end{linenomath*}
where $\sigma_{\text{T}}$ is the Thomson scattering cross-section for the electron ($6.65 \times 10^{-25} cm^2$), $D_\text{L}$ is the luminosity distance for the source ($20.85$ Gpc under standard cosmology parameters), $z$ is the redshift ($= 2.507$), $F_{\text{X-ray}}$ is the flux in the X-ray range ($0.3-8$ keV, $13.8 \times 10^{-12}$ erg $cm^{-2} s^{-1}$ for Flare A), $\varepsilon$ is the energy of the highest energy photon observed during the flare scaled by $m_e c^2$ ($24.3$ GeV for Flare A), $t_{\text{var}}$ is the variability time scale estimated in Section \ref{sec:doubling_time} and $m_e$ is the mass of the electron. This estimate assumes that the optical depth $\tau_{\gamma \gamma}$ is 1 for the highest energy photon. The $\delta_{\text{min}}$ comes out to be $32.8$ for Flare A and is very similar for rest of the flares since they have similar highest photon energies and X-ray flux.

\subsection{Broadband emission during flaring states}
Three flaring periods were identified between MJD 58572-58607 (30 March 2019 - 4 May 2019, Fig \ref{fig:combined_lightcurve}) and were modelled with a leptonic scenario. The peak flux during all the flares was $\sim 20 \times 10^{-6}$ photons cm$^{-2}$ s$^{-1}$ while it was below $\sim 20 \times 10^{-6}$ photons cm$^{-2}$ s$^{-1}$ during the pre-flare and post-flare states. Figure \ref{fig:sed_model_fit} shows the modelled SEDs of the flaring periods as well as pre-flare and post-flare states for comparison. The modelled parameters are mentioned in Table \ref{tab:sed_parameters}. 

Typical values were used for the energy density and temperature of the accretion disk in Table \ref{tab:sed_parameters}. The Energy density of the \gls{blr} was roughly estimated using Eq \ref{eq:blr_density} with $\Gamma = 35$, and $\eta_\text{BLR} \sim 0.05$ to get an  energy density of the order of 10 erg/cm$^{3}$ which was later fine tuned based on the fit to the \gls{sed} data points. A typical value was used for \gls{blr} temperature ($T_\text{BLR}$). The Doppler factor was roughly  estimated based on $\delta_{\text{min}} = 32.8$ discussed in Section \ref{sec:min_doppler} and increased slightly to $35$ to better fit the observations.

The model parameters during flares A, B and C were similar with only minor differences in the spectral indices, $\alpha$ ranging from $1.7$ to $1.8$ and $\beta$ changing from $0.3$ to $0.5$. The magnetic field strength was more or less constant at about $\sim 1.9$ Gauss. The magnetic field value $(B)$ was only constrained by the slope of the \gls{sed} in the UV-optical region since we have no radio band data, so there is some room for modification and improvement with observations in lower energy ranges. The pre-flare and post-flare states were found to be similar. The UV-optical emission of the source was dominated by the synchrotron emission rather than the thermal emission from the accretion disk, a trend also observed in our previous work on PKS 1424-418 \citep{2021MNRAS.501.2504A} and in 3C 279 \citep{2020ApJ...890..164P}. The X-ray and $\gamma$-ray flux were modelled to be originating from the \gls{ic} scattering of photons from the \gls{blr} based on the correlated X-ray and $\gamma$-ray flux and the location of the emission region found in Section \ref{sec:emission_region}. 

The flaring exhibited by the source during October 2010 was modelled by \cite{2015ApJ...799..143A} with a steady state \gls{ec}-\gls{dt} model, with notable differences between their model and this work being in our value of the Doppler factor ($\delta = 35$ compared to 20) and our magnetic field strength $B$ value being higher by a factor of $2$. This might be due to the fact that the $\gamma$-ray flux during April 2019 was an order of magnitude higher than the flux for the flaring observed in October 2010 in \cite{2015ApJ...799..143A} and the fact that we did not correct for dust extinction and magnification due to lensing. We also see a corresponding increase in the power of the jet in our model compared to \cite{2015ApJ...799..143A} while still being below the Eddington limit. We concluded that the flares studied in this work could be the result of increase Doppler factor, magnetic field and jet power. The increase in jet power can be associated with the increase in the accretion rate.

\subsection{Discussion on the mass of the central Black Hole}
The \gls{smbh} mass used in all the calculations in previous sections was $1 \times 10^9 M_\odot$ based on the mass used in \cite{2015ApJ...799..143A}. We could not trace the origin of this number in the literature on this source and suspect that the value is based on statistical estimates linking luminosity and \gls{smbh} mass for blazars. In the flares studied in this work, the flux and consequently the estimated jet power is much higher than previous studies and the jet power approaches the Eddington limit suggesting a very high Eddington ratio of $0.5-0.7$. Although this is not a problem as such, typically the values tend to be lower, ranging from $0.03-0.3$ \citep{10.1046/j.1365-8711.2003.06255.x}. We hence suggest a higher \gls{smbh} mass of $3-5 \times 10^9 M_\odot$, which is still within the \gls{fsrq} \gls{smbh} mass range (Fig 1 in \cite{10.1111/j.1365-2966.2008.13360.x}).

\section{Summary}
The source PKS 1830-211 was studied during its flaring state during the period Oct 2018 to Nov 2019. The $\gamma$-ray flux reached as high as $20 \times 10^{-6}$ photons cm$^{-2} s^{-1}$ during April 2019. The X-ray and $\gamma$-ray emission was found to be correlated with zero lag while no such correlation was observed between the $\gamma$-ray and UV-optical band emission. Gravitational lensing due to intervening galaxies was expected to lead to self-correlation in the $\gamma$-ray light curve at non-zero lag/lead based on previous studies but no such correlation was found. The X-ray emission from the source displayed a correlation between higher flux and lower spectral index while no such trend was observed for the $\gamma$-ray flux. We analysed the $\gamma$-ray light curve to estimate the variability, the rise and decay time scales were found to be of the order of 3 hours based on the lowest possible binning for the data with $t_{\text{var}} = 1.66$ hr. The highest energy photon detected was $24.3$ GeV which arrived during Flare A. 

We found that the $\gamma$-ray light curve did not have a strong and conclusive self-correlation at the expected time delay of about 26 days detected in radio bands by \cite{1996ApJ...472L...5L} and first reported for $\gamma$-rays by \cite{2011A&A...528L...3B}. A similar observation was also made in \cite{2015ApJ...799..143A} where the authors concluded that the flux ratio for the lensed images was lower for $\gamma$-rays compared to radio bands (1:6 instead of 1:1.5 found in \citealt{1996ApJ...472L...5L}) and only a small self-correlation peak was seen at 19 days instead of 26 days. While our analysis is not as rigorous as \cite{2011A&A...528L...3B} or \cite{2015ApJ...799..143A}, our self-correlation result is in line with \cite{2015ApJ...799..143A}. There can be multiple reasons for the absence of a strong self-correlation peak in the $\gamma$-ray light curve. The two lensed images, NE and SW are about $1''$ of each other \citep{1988MNRAS.231..229P} and hence susceptible to variation in the flux ratio of the images due to the anisotropy of relativistic beaming. Our analysis of the 2018-19 flaring episodes suggests a high bulk Lorentz factor of 35 which implies a narrower jet ($\theta \propto 1/\Gamma$; \citealt{Dermer_2009}) which can result in a high flux ratio between the lensed images since the intervening galaxies form an compound lens and the relativistic jet is likely to be asymmetric with respect to the lens.

Another possibility is that the milli/micro-lensing effects and consequently the flux ratio is frequency dependent (chromatic lensing, \cite{2006ApJ...640..569B, 2011ApJ...740L..34C, 2015ApJ...799..143A}) because of the differences in the size of the emission regions for the frequency bands. The emission region is much smaller for $\gamma$-rays compared to the radio bands. If the flux ratio is high, the signal can be lost in the intrinsic variability of the blazar which was found to be of the order of 3 hrs for PKS 1830-211 in this work, orders of magnitude smaller than the 26 day lensing delay. The variability timescales could be even lower if shorter time bins are used, if not for high uncertainty because of the low photon flux. \cite{2015ApJ...799..143A} also point to the possibility that the results in their work and \cite{2011A&A...528L...3B} are not necessarily mutually exclusive as the emission region might be different for the flares studied in the papers with different flux ratios, this explanation holds for our results as well.

The \gls{sed} was fitted using a leptonic model with a log-parabola injection spectrum with $\alpha \sim 1.8$, $\beta \sim 0.4$ and seed photons sourced from the \gls{blr} and the accretion disk. The size and the location of the emission region were found to be within the \gls{blr} and hence contribution from the \gls{dt} was not included. The UV-optical emission could be explained by synchrotron emission from the lepton population, the X-ray and $\gamma$-ray emission could be fitted with a single \gls{ec}-\gls{blr} component which supports the correlated flux found in the light curve of the two energy ranges. The three flaring periods A, B and C were found to have very similar model fit parameters and the same was the case for the pre-flare and post-flare periods. The power carried by the various components of the jet was estimated based on this model fit.

Our model uses a different set of components compared to \cite{2015ApJ...799..143A} where seed photons from the \gls{dt} were up-scattered to reproduce the \gls{sed}. \cite{2015ApJ...799..143A} used a steady-state broken power law distribution while our model is time dependent. Our magnetic field was a factor of 2 higher and consequently, the power carried by the Poynting flux was greater than their model. The spectral indices $\alpha$ were found to be around $1.8$ for all the flares and within the expected range for \glspl{fsrq}, similar to \cite{2015ApJ...799..143A} although the authors used a broken power law spectrum for the leptonic population. We modelled all the states with a Doppler factor of $35$ which is much higher than the value of $20$ used by \cite{2015ApJ...799..143A} although the flaring was also much stronger in our case and so was the $\delta_{\text{min}}$ as discussed in Section \ref{sec:min_doppler}. Our value of the minimum Lorentz factor for the leptons was also much higher at $9$ compared to $3$ in \cite{2015ApJ...799..143A} for the same reason. The maximum allowed Lorentz factor for leptons was much lower for our model since we have a log-parabola spectrum instead of the hard broken power law spectrum used in \cite{2015ApJ...799..143A}. Our broadband \gls{sed} modelling concludes that the increase in Doppler factor, magnetic field and high jet power might be responsible for the current highest flaring state of blazar PKS 1830-211.

\section*{Acknowledgements}
We thank the anonymous reviewer for useful and constructive comments that helped improve our manuscript. R. Prince is grateful for the support of the Polish Funding Agency National Science Centre, project 2017/26/-A/ST9/-00756 (MAESTRO 9) and MNiSW grant DIR/WK/2018/12. D. Bose acknowledges support of Ramanujan Fellowship- SB/S2/RJN-038/2017.

\twocolumn

\bibliographystyle{mnras}

\begin{thebibliography}{}
\expandafter\ifx\csname natexlab\endcsname\relax\def\natexlab#1{#1}\fi
\providecommand{\url}[1]{\href{#1}{#1}}
\providecommand{\dodoi}[1]{doi:~\href{http://doi.org/#1}{\nolinkurl{#1}}}
\providecommand{\doeprint}[1]{\href{http://ascl.net/#1}{\nolinkurl{http://ascl.net/#1}}}
\providecommand{\doarXiv}[1]{\href{https://arxiv.org/abs/#1}{\nolinkurl{https://arxiv.org/abs/#1}}}

\bibitem[{{Abdo} {et~al.}(2011){Abdo}, {Ackermann}, {Ajello}, {Allafort},
  {Baldini}, {Ballet}, {Barbiellini}, {Bastieri}, {Bellazzini}, {Berenji},
  {Blandford}, {Bloom}, {Bonamente}, {Borgland}, {Bouvier}, {Bregeon},
  {Brigida}, {Bruel}, {Buehler}, {Buson}, {Caliandro}, {Cameron}, {Caraveo},
  {Casandjian}, {Cavazzuti}, {Cecchi}, {Charles}, {Chekhtman}, {Cheung},
  {Chiang}, {Ciprini}, {Claus}, {Conrad}, {Cutini}, {D'Ammando}, {de Angelis},
  {de Palma}, {Dermer}, {Digel}, {Silva}, {Drell}, {Dubois}, {Dumora},
  {Escande}, {Favuzzi}, {Fegan}, {Ferrara}, {Fortin}, {Fukazawa}, {Fusco},
  {Gargano}, {Gasparrini}, {Gehrels}, {Germani}, {Giglietto}, {Giommi},
  {Giordano}, {Giroletti}, {Glanzman}, {Godfrey}, {Grenier}, {Grove},
  {Guiriec}, {Hadasch}, {Hayashida}, {Hays}, {Horan}, {Itoh},
  {J{\'o}hannesson}, {Johnson}, {Kamae}, {Katagiri}, {Kataoka},
  {Kn{\"o}dlseder}, {Kuss}, {Lande}, {Larsson}, {Latronico}, {Lee}, {Longo},
  {Loparco}, {Lott}, {Lovellette}, {Lubrano}, {Madejski}, {Makeev},
  {Mazziotta}, {McConville}, {McEnery}, {Michelson}, {Mitthumsiri}, {Mizuno},
  {Moiseev}, {Monte}, {Monzani}, {Morselli}, {Moskalenko}, {Murgia},
  {Naumann-Godo}, {Nishino}, {Nolan}, {Norris}, {Nuss}, {Ohsugi}, {Okumura},
  {Orlando}, {Ormes}, {Paneque}, {Pelassa}, {Pesce-Rollins}, {Pierbattista},
  {Piron}, {Porter}, {Rain{\`o}}, {Rando}, {Razzaque}, {Reimer}, {Reimer},
  {Ritz}, {Roth}, {Sadrozinski}, {Sanchez}, {Scargle}, {Schalk}, {Sgr{\`o}},
  {Siskind}, {Smith}, {Spandre}, {Spinelli}, {Strickman}, {Takahashi},
  {Takahashi}, {Tanaka}, {Tanaka}, {Thayer}, {Thayer}, {Thompson}, {Tibaldo},
  {Torres}, {Tosti}, {Tramacere}, {Troja}, {Vandenbroucke}, {Vasileiou},
  {Vianello}, {Vilchez}, {Vitale}, {Waite}, {Wang}, {Winer}, {Wood}, {Yang}, \&
  {Ziegler}}]{2011ApJ...733L..26A}
{Abdo}, A.~A., {Ackermann}, M., {Ajello}, M., {et~al.} 2011, \apjl, 733, L26,
  \dodoi{10.1088/2041-8205/733/2/L26}

\bibitem[{{Abdo} {et~al.}(2015){Abdo}, {Ackermann}, {Ajello}, {Allafort},
  {Amin}, {Baldini}, {Barbiellini}, {Bastieri}, {Bechtol}, {Bellazzini},
  {Blandford}, {Bonamente}, {Borgland}, {Bregeon}, {Brigida}, {Buehler},
  {Bulmash}, {Buson}, {Caliandro}, {Cameron}, {Caraveo}, {Cavazzuti}, {Cecchi},
  {Charles}, {Cheung}, {Chiang}, {Chiaro}, {Ciprini}, {Claus}, {Cohen-Tanugi},
  {Conrad}, {Corbet}, {Cutini}, {D'Ammando}, {de Angelis}, {de Palma},
  {Dermer}, {Drell}, {Drlica-Wagner}, {Favuzzi}, {Finke}, {Focke}, {Fukazawa},
  {Fusco}, {Gargano}, {Gasparrini}, {Gehrels}, {Giglietto}, {Giordano},
  {Giroletti}, {Glanzman}, {Grenier}, {Grove}, {Guiriec}, {Hadasch},
  {Hayashida}, {Hays}, {Hughes}, {Inoue}, {Jackson}, {Jogler},
  {J{\'o}hannesson}, {Johnson}, {Kamae}, {Kn{\"o}dlseder}, {Kuss}, {Lande},
  {Larsson}, {Latronico}, {Longo}, {Loparco}, {Lott}, {Lovellette}, {Lubrano},
  {Madejski}, {Mazziotta}, {Mehault}, {Michelson}, {Mizuno}, {Monzani},
  {Morselli}, {Moskalenko}, {Murgia}, {Nemmen}, {Nuss}, {Ohno}, {Ohsugi},
  {Paneque}, {Perkins}, {Pesce-Rollins}, {Piron}, {Pivato}, {Porter},
  {Rain{\`o}}, {Rando}, {Razzano}, {Reimer}, {Reimer}, {Reyes}, {Ritz},
  {Romoli}, {Roth}, {Saz Parkinson}, {Sgr{\`o}}, {Siskind}, {Spandre},
  {Spinelli}, {Takahashi}, {Takeuchi}, {Tanaka}, {Thayer}, {Thayer},
  {Thompson}, {Tibaldo}, {Tinivella}, {Torres}, {Tosti}, {Troja}, {Tronconi},
  {Usher}, {Vandenbroucke}, {Vasileiou}, {Vianello}, {Vitale}, {Waite},
  {Werner}, {Winer}, \& {Wood}}]{2015ApJ...799..143A}
---. 2015, \apj, 799, 143, \dodoi{10.1088/0004-637X/799/2/143}

\bibitem[{{Abhir} {et~al.}(2021){Abhir}, {Joseph}, {Patel}, \&
  {Bose}}]{2021MNRAS.501.2504A}
{Abhir}, J., {Joseph}, J., {Patel}, S.~R., \& {Bose}, D. 2021, \mnras, 501,
  2504, \dodoi{10.1093/mnras/staa3639}

\bibitem[{{Acero} {et~al.}(2015){Acero}, {Ackermann}, {Ajello}, {Albert},
  {Atwood}, {Axelsson}, {Baldini}, {Ballet}, {Barbiellini}, {Bastieri},
  {Belfiore}, {Bellazzini}, {Bissaldi}, {Blandford}, {Bloom}, {Bogart},
  {Bonino}, {Bottacini}, {Bregeon}, {Britto}, {Bruel}, {Buehler}, {Burnett},
  {Buson}, {Caliandro}, {Cameron}, {Caputo}, {Caragiulo}, {Caraveo},
  {Casandjian}, {Cavazzuti}, {Charles}, {Chaves}, {Chekhtman}, {Cheung},
  {Chiang}, {Chiaro}, {Ciprini}, {Claus}, {Cohen-Tanugi}, {Cominsky}, {Conrad},
  {Cutini}, {D'Ammando}, {de Angelis}, {DeKlotz}, {de Palma}, {Desiante},
  {Digel}, {Di Venere}, {Drell}, {Dubois}, {Dumora}, {Favuzzi}, {Fegan},
  {Ferrara}, {Finke}, {Franckowiak}, {Fukazawa}, {Funk}, {Fusco}, {Gargano},
  {Gasparrini}, {Giebels}, {Giglietto}, {Giommi}, {Giordano}, {Giroletti},
  {Glanzman}, {Godfrey}, {Grenier}, {Grondin}, {Grove}, {Guillemot}, {Guiriec},
  {Hadasch}, {Harding}, {Hays}, {Hewitt}, {Hill}, {Horan}, {Iafrate}, {Jogler},
  {J{\'o}hannesson}, {Johnson}, {Johnson}, {Johnson}, {Johnson}, {Kamae},
  {Kataoka}, {Katsuta}, {Kuss}, {La Mura}, {Landriu}, {Larsson}, {Latronico},
  {Lemoine-Goumard}, {Li}, {Li}, {Longo}, {Loparco}, {Lott}, {Lovellette},
  {Lubrano}, {Madejski}, {Massaro}, {Mayer}, {Mazziotta}, {McEnery},
  {Michelson}, {Mirabal}, {Mizuno}, {Moiseev}, {Mongelli}, {Monzani},
  {Morselli}, {Moskalenko}, {Murgia}, {Nuss}, {Ohno}, {Ohsugi}, {Omodei},
  {Orienti}, {Orlando}, {Ormes}, {Paneque}, {Panetta}, {Perkins},
  {Pesce-Rollins}, {Piron}, {Pivato}, {Porter}, {Racusin}, {Rando}, {Razzano},
  {Razzaque}, {Reimer}, {Reimer}, {Reposeur}, {Rochester}, {Romani},
  {Salvetti}, {S{\'a}nchez-Conde}, {Saz Parkinson}, {Schulz}, {Siskind},
  {Smith}, {Spada}, {Spandre}, {Spinelli}, {Stephens}, {Strong}, {Suson},
  {Takahashi}, {Takahashi}, {Tanaka}, {Thayer}, {Thayer}, {Thompson},
  {Tibaldo}, {Tibolla}, {Torres}, {Torresi}, {Tosti}, {Troja}, {Van Klaveren},
  {Vianello}, {Winer}, {Wood}, {Wood}, {Zimmer}, \& {Fermi-LAT
  Collaboration}}]{3fgl_catalog}
{Acero}, F., {Ackermann}, M., {Ajello}, M., {et~al.} 2015, \apjs, 218, 23,
  \dodoi{10.1088/0067-0049/218/2/23}

\bibitem[{Ajello {et~al.}(2020)Ajello, Angioni, Axelsson, Ballet, Barbiellini,
  Bastieri, Gonzalez, Bellazzini, Bissaldi, Bloom, Bonino, Bottacini, Bruel,
  Buson, Cafardo, Cameron, Cavazzuti, Chen, Cheung, Ciprini, Costantin, Cutini,
  D'Ammando, de~la Torre~Luque, de~Menezes, de~Palma, Desai, Lalla, Venere,
  Dom{\'{\i}}nguez, Dirirsa, Ferrara, Finke, Franckowiak, Fukazawa, Funk,
  Fusco, Gargano, Garrappa, Gasparrini, Giglietto, Giordano, Giroletti, Green,
  Grenier, Guiriec, Harita, Hays, Horan, Itoh, J{\'{o}}hannesson, Kovac'evic',
  Krauss, Kreter, Kuss, Larsson, Leto, Li, Liodakis, Longo, Loparco, Lott,
  Lovellette, Lubrano, Madejski, Maldera, Manfreda, Mart{\'{\i}}-Devesa,
  Massaro, Mazziotta, Mereu, Meyer, Migliori, Mirabal, Mizuno, Monzani,
  Morselli, Moskalenko, Negro, Nemmen, Nuss, Ojha, Ojha, Omodei, Orienti,
  Orlando, Ormes, Paliya, Pei, Pe{\~{n}}a-Herazo, Persic, Pesce-Rollins,
  Petrov, Piron, Poon, Principe, Rain{\`{o}}, Rando, Rani, Razzano, Razzaque,
  Reimer, Reimer, Schinzel, Serini, Sgr{\`{o}}, Siskind, Spandre, Spinelli,
  Suson, Tachibana, Thompson, Torres, Torresi, Troja, Valverde, van Zyl, \&
  Yassine}]{Ajello_2020}
Ajello, M., Angioni, R., Axelsson, M., {et~al.} 2020, The Astrophysical
  Journal, 892, 105, \dodoi{10.3847/1538-4357/ab791e}

\bibitem[{{Alexander}(2013)}]{2013arXiv1302.1508A}
{Alexander}, T. 2013, arXiv e-prints, arXiv:1302.1508.
\newblock \doarXiv{1302.1508}

\bibitem[{{Arnaud}(1996)}]{1996ASPC..101...17A}
{Arnaud}, K.~A. 1996, in Astronomical Society of the Pacific Conference Series,
  Vol. 101, Astronomical Data Analysis Software and Systems V, ed. G.~H.
  {Jacoby} \& J.~{Barnes}, 17

\bibitem[{{Atwood} {et~al.}(2009){Atwood}, {Abdo}, {Ackermann}, {Althouse},
  {Anderson}, {Axelsson}, {Baldini}, {Ballet}, {Band}, {Barbiellini},
  {Bartelt}, {Bastieri}, {Baughman}, {Bechtol}, {B{\'e}d{\'e}r{\`e}de},
  {Bellardi}, {Bellazzini}, {Berenji}, {Bignami}, {Bisello}, {Bissaldi},
  {Blandford}, {Bloom}, {Bogart}, {Bonamente}, {Bonnell}, {Borgland },
  {Bouvier}, {Bregeon}, {Brez}, {Brigida}, {Bruel}, {Burnett}, {Busetto},
  {Caliandro}, {Cameron}, {Caraveo}, {Carius}, {Carlson}, {Casandjian},
  {Cavazzuti}, {Ceccanti}, {Cecchi}, {Charles}, {Chekhtman}, {Cheung},
  {Chiang}, {Chipaux}, {Cillis}, {Ciprini}, {Claus}, {Cohen-Tanugi},
  {Condamoor}, {Conrad}, {Corbet}, {Corucci}, {Costamante}, {Cutini}, {Davis},
  {Decotigny}, {DeKlotz}, {Dermer}, {de Angelis}, {Digel}, {do Couto e Silva},
  {Drell}, {Dubois}, {Dumora}, {Edmonds}, {Fabiani}, {Farnier}, {Favuzzi},
  {Flath}, {Fleury}, {Focke}, {Funk}, {Fusco}, {Gargano}, {Gasparrini},
  {Gehrels}, {Gentit}, {Germani}, {Giebels}, {Giglietto}, {Giommi}, {Giordano},
  {Glanzman}, {Godfrey}, {Grenier}, {Grondin}, {Grove}, {Guillemot}, {Guiriec},
  {Haller}, {Harding}, {Hart}, {Hays}, {Healey}, {Hirayama}, {Hjalmarsdotter},
  {Horn}, {Hughes}, {J{\'o}hannesson}, {Johansson}, {Johnson}, {Johnson},
  {Johnson}, {Johnson}, {Kamae}, {Katagiri}, {Kataoka}, {Kavelaars}, {Kawai},
  {Kelly}, {Kerr}, {Klamra}, {Kn{\"o}dlseder}, {Kocian}, {Komin}, {Kuehn},
  {Kuss}, {Landriu}, {Latronico}, {Lee}, {Lee}, {Lemoine-Goumard}, {Lionetto},
  {Longo}, {Loparco}, {Lott}, {Lovellette}, {Lubrano}, {Madejski}, {Makeev},
  {Marangelli}, {Massai}, {Mazziotta}, {McEnery}, {Menon}, {Meurer},
  {Michelson}, {Minuti}, {Mirizzi}, {Mitthumsiri}, {Mizuno}, {Moiseev},
  {Monte}, {Monzani}, {Moretti}, {Morselli}, {Moskalenko}, {Murgia},
  {Nakamori}, {Nishino}, {Nolan}, {Norris}, {Nuss}, {Ohno}, {Ohsugi}, {Omodei},
  {Orlando}, {Ormes}, {Paccagnella}, {Paneque}, {Panetta}, {Parent}, {Pearce},
  {Pepe}, {Perazzo}, {Pesce-Rollins}, {Picozza}, {Pieri}, {Pinchera}, {Piron},
  {Porter}, {Poupard}, {Rain{\`o}}, {Rando}, {Rapposelli}, {Razzano}, {Reimer},
  {Reimer}, {Reposeur}, {Reyes}, {Ritz}, {Rochester}, {Rodriguez}, {Romani},
  {Roth}, {Russell}, {Ryde}, {Sabatini}, {Sadrozinski}, {Sanchez}, {Sand er},
  {Sapozhnikov}, {Parkinson}, {Scargle}, {Schalk}, {Scolieri}, {Sgr{\`o}},
  {Share}, {Shaw}, {Shimokawabe}, {Shrader}, {Sierpowska-Bartosik}, {Siskind},
  {Smith}, {Smith}, {Spandre}, {Spinelli}, {Starck}, {Stephens}, {Strickman},
  {Strong}, {Suson}, {Tajima}, {Takahashi}, {Takahashi}, {Tanaka}, {Tenze},
  {Tether}, {Thayer}, {Thayer}, {Thompson}, {Tibaldo}, {Tibolla}, {Torres},
  {Tosti}, {Tramacere}, {Turri}, {Usher}, {Vilchez}, {Vitale}, {Wang},
  {Watters}, {Winer}, {Wood}, {Ylinen}, \& {Ziegler}}]{2009ApJ...697.1071A}
{Atwood}, W.~B., {Abdo}, A.~A., {Ackermann}, M., {et~al.} 2009, \apj, 697,
  1071, \dodoi{10.1088/0004-637X/697/2/1071}

\bibitem[{{Barnacka} {et~al.}(2011){Barnacka}, {Glicenstein}, \&
  {Moudden}}]{2011A&A...528L...3B}
{Barnacka}, A., {Glicenstein}, J.~F., \& {Moudden}, Y. 2011, \aap, 528, L3,
  \dodoi{10.1051/0004-6361/201016175}

\bibitem[{{Blackburne} {et~al.}(2006){Blackburne}, {Pooley}, \&
  {Rappaport}}]{2006ApJ...640..569B}
{Blackburne}, J.~A., {Pooley}, D., \& {Rappaport}, S. 2006, \apj, 640, 569,
  \dodoi{10.1086/500172}

\bibitem[{Breeveld {et~al.}(2011)Breeveld, Landsman, Holland, Roming, Kuin, \&
  Page}]{Breeveld_2011}
Breeveld, A.~A., Landsman, W., Holland, S.~T., {et~al.} 2011, AIP Conference
  Proceedings, 1358, 373, \dodoi{10.1063/1.3621807}

\bibitem[{{Burrows} {et~al.}(2004){Burrows}, {Hill}, {Nousek}, {Wells},
  {Chincarini}, {Abbey}, {Beardmore}, {Bosworth}, {Br{\"a}uninger}, {Burkert},
  {Campana}, {Capalbi}, {Chang}, {Citterio}, {Freyberg}, {Giommi}, {Hartner},
  {Killough}, {Kittle}, {Klar}, {Mangels}, {McMeekin}, {Miles}, {Moretti},
  {Mori}, {Morris}, {Mukerjee}, {Osborne}, {Short}, {Tagliaferri},
  {Tamburelli}, {Watson}, {Willingale}, \& {Zugger}}]{2004SPIE.5165..201B}
{Burrows}, D.~N., {Hill}, J.~E., {Nousek}, J.~A., {et~al.} 2004, in Society of
  Photo-Optical Instrumentation Engineers (SPIE) Conference Series, Vol. 5165,
  X-Ray and Gamma-Ray Instrumentation for Astronomy XIII, ed. K.~A. {Flanagan}
  \& O.~H.~W. {Siegmund}, 201--216, \dodoi{10.1117/12.504868}

\bibitem[{{Cash}(1979)}]{1979ApJ...228..939C}
{Cash}, W. 1979, \apj, 228, 939, \dodoi{10.1086/156922}

\bibitem[{Celotti \& Ghisellini(2008)}]{Celotti2008}
Celotti, A., \& Ghisellini, G. 2008, Monthly Notices of the Royal Astronomical
  Society, 385, 283, \dodoi{10.1111/j.1365-2966.2007.12758.x}

\bibitem[{{Chen} {et~al.}(2011){Chen}, {Dai}, {Kochanek}, {Chartas},
  {Blackburne}, \& {Koz{\l}owski}}]{2011ApJ...740L..34C}
{Chen}, B., {Dai}, X., {Kochanek}, C.~S., {et~al.} 2011, \apjl, 740, L34,
  \dodoi{10.1088/2041-8205/740/2/L34}

\bibitem[{{Cheung} {et~al.}(2014){Cheung}, {Larsson}, {Scargle}, {Amin},
  {Blandford}, {Bulmash}, {Chiang}, {Ciprini}, {Corbet}, {Falco}, {Marshall},
  {Wood}, {Ajello}, {Bastieri}, {Chekhtman}, {D'Ammando}, {Giroletti}, {Grove},
  {Lott}, {Ojha}, {Orienti}, {Perkins}, {Razzano}, {Smith}, {Thompson}, \&
  {Wood}}]{2014ApJ...782L..14C}
{Cheung}, C.~C., {Larsson}, S., {Scargle}, J.~D., {et~al.} 2014, \apjl, 782,
  L14, \dodoi{10.1088/2041-8205/782/2/L14}

\bibitem[{D'Elia {et~al.}(2003)D'Elia, Padovani, \&
  Landt}]{10.1046/j.1365-8711.2003.06255.x}
D'Elia, V., Padovani, P., \& Landt, H. 2003, Monthly Notices of the Royal
  Astronomical Society, 339, 1081, \dodoi{10.1046/j.1365-8711.2003.06255.x}

\bibitem[{{Dermer} \& {Menon}(2009)}]{Dermer_2009}
{Dermer}, C.~D., \& {Menon}, G. 2009, {High Energy Radiation from Black Holes:
  Gamma Rays, Cosmic Rays, and Neutrinos} (Princeton University Press)

\bibitem[{{Dermer} \& {Schlickeiser}(1993)}]{1993ApJ...416..458D}
{Dermer}, C.~D., \& {Schlickeiser}, R. 1993, \apj, 416, 458,
  \dodoi{10.1086/173251}

\bibitem[{{Edelson} \& {Krolik}(1988)}]{1988ApJ...333..646E}
{Edelson}, R.~A., \& {Krolik}, J.~H. 1988, \apj, 333, 646,
  \dodoi{10.1086/166773}

\bibitem[{{Foschini} {et~al.}(2006){Foschini}, {Ghisellini}, {Raiteri},
  {Tavecchio}, {Villata}, {Maraschi}, {Pian}, {Tagliaferri}, {Di Cocco}, \&
  {Malaguti}}]{2006A&A...453..829F}
{Foschini}, L., {Ghisellini}, G., {Raiteri}, C.~M., {et~al.} 2006, \aap, 453,
  829, \dodoi{10.1051/0004-6361:20064921}

\bibitem[{{Franceschini} {et~al.}(2008){Franceschini}, {Rodighiero}, \&
  {Vaccari}}]{2008A&A...487..837F}
{Franceschini}, A., {Rodighiero}, G., \& {Vaccari}, M. 2008, \aap, 487, 837,
  \dodoi{10.1051/0004-6361:200809691}

\bibitem[{{Ghisellini} {et~al.}(1985){Ghisellini}, {Maraschi}, \&
  {Treves}}]{1985A&A...146..204G}
{Ghisellini}, G., {Maraschi}, L., \& {Treves}, A. 1985, \aap, 146, 204

\bibitem[{Ghisellini \& Tavecchio(2008)}]{10.1111/j.1365-2966.2008.13360.x}
Ghisellini, G., \& Tavecchio, F. 2008, Monthly Notices of the Royal
  Astronomical Society, 387, 1669, \dodoi{10.1111/j.1365-2966.2008.13360.x}

\bibitem[{Ghisellini \& Tavecchio(2009)}]{Ghisellini2009}
---. 2009, Monthly Notices of the Royal Astronomical Society, 397, 985,
  \dodoi{10.1111/j.1365-2966.2009.15007.x}

\bibitem[{{Giommi} {et~al.}(2006){Giommi}, {Blustin}, {Capalbi1},
  {Colafrancesco}, {Cucchiara}, {Fuhrmann}, {Krimm}, {Marchili}, {Massaro},
  {Perri}, {Tagliaferri}.G., {Tosti}, {Tramacere}, {Burrows}, {Chincarini},
  {Falcone}, {Gehrels}, {Kennea}, \& {Sambruna}}]{Giommi2006}
{Giommi}, P., {Blustin}, A., {Capalbi1}, M., {et~al.} 2006, \aap, 456, 911,
  \dodoi{10.1051/0004-6361:20064874}

\bibitem[{{Goyal}(2020)}]{Goyal_2020}
{Goyal}, A. 2020, \mnras, 494, 3432, \dodoi{10.1093/mnras/staa997}

\bibitem[{{Goyal} {et~al.}(2017){Goyal}, {Stawarz}, {Ostrowski}, {Larionov},
  {Gopal-Krishna}, {Wiita}, {Joshi}, {Soida}, \& {Agudo}}]{Goyal_2017}
{Goyal}, A., {Stawarz}, {\L}., {Ostrowski}, M., {et~al.} 2017, \apj, 837, 127,
  \dodoi{10.3847/1538-4357/aa6000}

\bibitem[{{Goyal} {et~al.}(2018){Goyal}, {Stawarz}, {Zola}, {Marchenko},
  {Soida}, {Nilsson}, {Ciprini}, {Baran}, {Ostrowski}, {Wiita},
  {Gopal-Krishna}, {Siemiginowska}, {Sobolewska}, {Jorstad}, {Marscher},
  {Aller}, {Aller}, {Hovatta}, {Caton}, {Reichart}, {Matsumoto}, {Sadakane},
  {Gazeas}, {Kidger}, {Piirola}, {Jermak}, {Alicavus}, {Baliyan}, {Baransky},
  {Berdyugin}, {Blay}, {Boumis}, {Boyd}, {Bufan}, {Campas Torrent}, {Campos},
  {Carrillo G{\'o}mez}, {Dalessio}, {Debski}, {Dimitrov}, {Drozdz}, {Er},
  {Erdem}, {Escartin P{\'e}rez}, {Fallah Ramazani}, {Filippenko}, {Gafton},
  {Garcia}, {Godunova}, {G{\'o}mez Pinilla}, {Gopinathan}, {Haislip}, {Haque},
  {Harmanen}, {Hudec}, {Hurst}, {Ivarsen}, {Joshi}, {Kagitani}, {Karaman},
  {Karjalainen}, {Kaur}, {Kozie{\l}-Wierzbowska}, {Kuligowska}, {Kundera},
  {Kurowski}, {Kvammen}, {LaCluyze}, {Lee}, {Liakos}, {Lozano de Haro},
  {Moore}, {Mugrauer}, {Naves Nogues}, {Neely}, {Ogloza}, {Okano}, {Pajdosz},
  {Pandey}, {Perri}, {Poyner}, {Provencal}, {Pursimo}, {Raj}, {Rajkumar},
  {Reinthal}, {Reynolds}, {Saario}, {Sadegi}, {Sakanoi}, {Salto Gonz{\'a}lez},
  {Sameer}, {Simon}, {Siwak}, {Schweyer}, {Sold{\'a}n Alfaro}, {Sonbas},
  {Strobl}, {Takalo}, {Tremosa Espasa}, {Valdes}, {Vasylenko}, {Verrecchia},
  {Webb}, {Yoneda}, {Zejmo}, {Zheng}, {Zielinski}, {Janik}, {Chavushyan},
  {Mohammed}, {Cheung}, \& {Giroletti}}]{Goyal_2018}
{Goyal}, A., {Stawarz}, {\L}., {Zola}, S., {et~al.} 2018, \apj, 863, 175,
  \dodoi{10.3847/1538-4357/aad2de}

\bibitem[{{H.~E.~S.~S. Collaboration} {et~al.}(2019){H.~E.~S.~S.
  Collaboration}, {Abdalla}, {Aharonian}, {Ait Benkhali}, {Ang{\"u}ner},
  {Arakawa}, {Arcaro}, {Armand}, {Arrieta}, {Backes}, {Barnard}, {Becherini},
  {Becker Tjus}, {Berge}, {Bernl{\"o}hr}, {Blackwell}, {B{\"o}ttcher},
  {Boisson}, {Bolmont}, {Bonnefoy}, {Bordas}, {Bregeon}, {Brun}, {Brun},
  {Bryan}, {B{\"u}chele}, {Bulik}, {Bylund}, {Capasso}, {Caroff}, {Carosi},
  {Casanova}, {Cerruti}, {Chakraborty}, {Chand}, {Chandra}, {Chaves}, {Chen},
  {Colafrancesco}, {Condon}, {Davids}, {Deil}, {Devin}, {deWilt}, {Dirson},
  {Djannati-Ata{\"\i}}, {Dmytriiev}, {Donath}, {Doroshenko}, {Drury}, {Dyks},
  {Egberts}, {Emery}, {Ernenwein}, {Eschbach}, {Fegan}, {Fiasson}, {Fontaine},
  {Funk}, {F{\"u}{\ss}ling}, {Gabici}, {Gallant}, {Gat{\'e}}, {Giavitto},
  {Glawion}, {Glicenstein}, {Gottschall}, {Grondin}, {Hahn}, {Haupt},
  {Heinzelmann}, {Henri}, {Hermann}, {Hinton}, {Hofmann}, {Hoischen}, {Holch},
  {Holler}, {Horns}, {Huber}, {Iwasaki}, {Jacholkowska}, {Jamrozy},
  {Jankowsky}, {Jankowsky}, {Jouvin}, {Jung-Richardt}, {Kastendieck},
  {Katarzy{\'n}ski}, {Katsuragawa}, {Katz}, {Khangulyan}, {Kh{\'e}lifi},
  {King}, {Klepser}, {Klu{\'z}niak}, {Komin}, {Kosack}, {Kraus}, {Lamanna},
  {Lau}, {Lefaucheur}, {Lemi{\`e}re}, {Lemoine-Goumard}, {Lenain}, {Leser},
  {Lohse}, {L{\'o}pez-Coto}, {Lorentz}, {Lypova}, {Malyshev}, {Marandon},
  {Marcowith}, {Mariaud}, {Mart{\'\i}-Devesa}, {Marx}, {Maurin}, {Meintjes},
  {Mitchell}, {Moderski}, {Mohamed}, {Mohrmann}, {Moore}, {Moulin}, {Murach},
  {Nakashima}, {de Naurois}, {Ndiyavala}, {Niederwanger}, {Niemiec}, {Oakes},
  {O'Brien}, {Odaka}, {Ohm}, {Ostrowski}, {Oya}, {Panter}, {Parsons},
  {Perennes}, {Petrucci}, {Peyaud}, {Piel}, {Pita}, {Poireau}, {Priyana Noel},
  {Prokhorov}, {Prokoph}, {P{\"u}hlhofer}, {Punch}, {Quirrenbach}, {Raab},
  {Rauth}, {Reimer}, {Reimer}, {Renaud}, {Rieger}, {Rinchiuso}, {Romoli},
  {Rowell}, {Rudak}, {Ruiz-Velasco}, {Sahakian}, {Saito}, {Sanchez},
  {Santangelo}, {Sasaki}, {Schlickeiser}, {Sch{\"u}ssler}, {Schulz}, {Schutte},
  {Schwanke}, {Schwemmer}, {Seglar-Arroyo}, {Senniappan}, {Seyffert}, {Shafi},
  {Shilon}, {Shiningayamwe}, {Simoni}, {Sinha}, {Sol}, {Specovius},
  {Spir-Jacob}, {Stawarz}, {Steenkamp}, {Stegmann}, {Steppa}, {Takahashi},
  {Tavernet}, {Tavernier}, {Taylor}, {Terrier}, {Tiziani}, {Tluczykont},
  {Trichard}, {Tsirou}, {Tsuji}, {Tuffs}, {Uchiyama}, {van der Walt}, {van
  Eldik}, {van Rensburg}, {van Soelen}, {Vasileiadis}, {Veh}, {Venter},
  {Vincent}, {Vink}, {Voisin}, {V{\"o}lk}, {Vuillaume}, {Wadiasingh}, {Wagner},
  {Wagner}, {White}, {Wierzcholska}, {Yang}, {Yoneda}, {Zaborov}, {Zacharias},
  {Zanin}, {Zdziarski}, {Zech}, {Ziegler}, {Zorn}, \&
  {{\.Z}ywucka}}]{2019MNRAS.486.3886H}
{H.~E.~S.~S. Collaboration}, {Abdalla}, H., {Aharonian}, F., {et~al.} 2019,
  \mnras, 486, 3886, \dodoi{10.1093/mnras/stz1031}

\bibitem[{{Hahn}(2015)}]{2015ICRC...34..917H}
{Hahn}, J. 2015, in International Cosmic Ray Conference, Vol.~34, 34th
  International Cosmic Ray Conference (ICRC2015), 917

\bibitem[{{HI4PI Collaboration} {et~al.}(2016){HI4PI Collaboration}, {Ben
  Bekhti}, {Fl{\"o}er}, {Keller}, {Kerp}, {Lenz}, {Winkel}, {Bailin},
  {Calabretta}, {Dedes}, {Ford}, {Gibson}, {Haud}, {Janowiecki}, {Kalberla},
  {Lockman}, {McClure-Griffiths}, {Murphy}, {Nakanishi}, {Pisano}, \&
  {Staveley-Smith}}]{2016A&A...594A.116H}
{HI4PI Collaboration}, {Ben Bekhti}, N., {Fl{\"o}er}, L., {et~al.} 2016, \aap,
  594, A116, \dodoi{10.1051/0004-6361/201629178}

\bibitem[{{Lidman} {et~al.}(1999){Lidman}, {Courbin}, {Meylan}, {Broadhurst},
  {Frye}, \& {Welch}}]{1999ApJ...514L..57L}
{Lidman}, C., {Courbin}, F., {Meylan}, G., {et~al.} 1999, \apjl, 514, L57,
  \dodoi{10.1086/311949}

\bibitem[{{Lovell} {et~al.}(1996){Lovell}, {Reynolds}, {Jauncey}, {Backus},
  {McCulloch}, {Sinclair}, {Wilson}, {Tzioumis}, {King}, {Gough}, {Ellingsen},
  {Phillips}, {Preston}, \& {Jones}}]{1996ApJ...472L...5L}
{Lovell}, J.~E.~J., {Reynolds}, J.~E., {Jauncey}, D.~L., {et~al.} 1996, \apjl,
  472, L5, \dodoi{10.1086/310353}

\bibitem[{{Maraschi} {et~al.}(1992){Maraschi}, {Ghisellini}, \&
  {Celotti}}]{1992ApJ...397L...5M}
{Maraschi}, L., {Ghisellini}, G., \& {Celotti}, A. 1992, \apjl, 397, L5,
  \dodoi{10.1086/186531}

\bibitem[{{Mattox} {et~al.}(1996){Mattox}, {Bertsch}, {Chiang}, {Dingus},
  {Digel}, {Esposito}, {Fierro}, {Hartman}, {Hunter}, {Kanbach}, {Kniffen},
  {Lin}, {Macomb}, {Mayer-Hasselwander}, {Michelson}, {von Montigny},
  {Mukherjee}, {Nolan}, {Ramanamurthy}, {Schneid}, {Sreekumar}, {Thompson}, \&
  {Willis}}]{1996ApJ...461..396M}
{Mattox}, J.~R., {Bertsch}, D.~L., {Chiang}, J., {et~al.} 1996, \apj, 461, 396,
  \dodoi{10.1086/177068}

\bibitem[{{Muller} {et~al.}(2020){Muller}, {Jaswanth}, {Horellou}, \&
  {Mart{\'\i}-Vidal}}]{2020A&A...641L...2M}
{Muller}, S., {Jaswanth}, S., {Horellou}, C., \& {Mart{\'\i}-Vidal}, I. 2020,
  \aap, 641, L2, \dodoi{10.1051/0004-6361/202038978}

\bibitem[{Paliya {et~al.}(2016)Paliya, Diltz, Böttcher, Stalin, \&
  Buckley}]{Paliya_2016}
Paliya, V.~S., Diltz, C., Böttcher, M., Stalin, C.~S., \& Buckley, D. 2016,
  The Astrophysical Journal, 817, 61, \dodoi{10.3847/0004-637x/817/1/61}

\bibitem[{Paliya {et~al.}(2015)Paliya, Sahayanathan, \& Stalin}]{Paliya_2015}
Paliya, V.~S., Sahayanathan, S., \& Stalin, C.~S. 2015, The Astrophysical
  Journal, 803, 15, \dodoi{10.1088/0004-637x/803/1/15}

\bibitem[{{Prince}(2020{\natexlab{a}})}]{Prince_2020}
{Prince}, R. 2020{\natexlab{a}}, \apj, 890, 164,
  \dodoi{10.3847/1538-4357/ab6b1e}

\bibitem[{{Prince}(2020{\natexlab{b}})}]{2020ApJ...890..164P}
---. 2020{\natexlab{b}}, \apj, 890, 164, \dodoi{10.3847/1538-4357/ab6b1e}

\bibitem[{{Rao} \& {Subrahmanyan}(1988)}]{1988MNRAS.231..229P}
{Rao}, A.~P., \& {Subrahmanyan}, R. 1988, \mnras, 231, 229,
  \dodoi{10.1093/mnras/231.2.229}

\bibitem[{{Roming} {et~al.}(2005){Roming}, {Kennedy}, {Mason}, {Nousek}, {Ahr},
  {Bingham}, {Broos}, {Carter}, {Hancock}, {Huckle}, {Hunsberger}, {Kawakami},
  {Killough}, {Koch}, {McLelland}, {Smith}, {Smith}, {Soto}, {Boyd},
  {Breeveld}, {Holland}, {Ivanushkina}, {Pryzby}, {Still}, \&
  {Stock}}]{Roming2005}
{Roming}, P.~W.~A., {Kennedy}, T.~E., {Mason}, K.~O., {et~al.} 2005, \ssr, 120,
  95, \dodoi{10.1007/s11214-005-5095-4}

\bibitem[{Schlafly \& Finkbeiner(2011)}]{Schlafly_2011}
Schlafly, E.~F., \& Finkbeiner, D.~P. 2011, The Astrophysical Journal, 737,
  103, \dodoi{10.1088/0004-637x/737/2/103}

\bibitem[{{Sikora} {et~al.}(1994){Sikora}, {Begelman}, \&
  {Rees}}]{1994ApJ...421..153S}
{Sikora}, M., {Begelman}, M.~C., \& {Rees}, M.~J. 1994, \apj, 421, 153,
  \dodoi{10.1086/173633}

\bibitem[{{Tarnopolski} {et~al.}(2020){Tarnopolski}, {{\.Z}ywucka},
  {Marchenko}, \& {Pascual-Granado}}]{2020ApJS..250....1T}
{Tarnopolski}, M., {{\.Z}ywucka}, N., {Marchenko}, V., \& {Pascual-Granado}, J.
  2020, \apjs, 250, 1, \dodoi{10.3847/1538-4365/aba2c7}

\bibitem[{Urry \& Padovani(1995)}]{Urry_1995}
Urry, C.~M., \& Padovani, P. 1995, Publications of the Astronomical Society of
  the Pacific, 107, 803, \dodoi{10.1086/133630}

\bibitem[{{Wood} {et~al.}(2017){Wood}, {Caputo}, {Charles}, {Di Mauro},
  {Magill}, {Perkins}, \& {Fermi-LAT Collaboration}}]{2017ICRC...35..824W}
{Wood}, M., {Caputo}, R., {Charles}, E., {et~al.} 2017, in International Cosmic
  Ray Conference, Vol. 301, 35th International Cosmic Ray Conference
  (ICRC2017), 824.
\newblock \doarXiv{1707.09551}

\end{thebibliography}

\printglossary[title=ABBREVIATIONS, nonumberlist]

\end{document}